\documentclass[11pt]{article}
\usepackage{amsmath,amssymb,color,slashed}
\usepackage{amsfonts}

\newcommand{\hoch}[1]{$\, ^{#1}$}

\newcommand{\be}{\begin{equation}}
\newcommand{\ee}{\end{equation}}
\newcommand{\bea}{\setlength\arraycolsep{2pt} \begin{eqnarray}}
\newcommand{\eea}{\end{eqnarray}}

\def\ft#1#2{{\textstyle{\frac{\scriptstyle #1}{\scriptstyle #2} } }}
\def\fft#1#2{{\frac{#1}{#2}}}
\def\CP{{{\mathbb C}{\mathbb P}}}
\def\0{{\sst{(0)}}}
\def\1{{\sst{(1)}}}
\def\2{{\sst{(2)}}}
\def\3{{\sst{(3)}}}
\def\4{{\sst{(4)}}}
\def\5{{\sst{(5)}}}
\def\6{{\sst{(6)}}}
\def\7{{\sst{(7)}}}
\def\8{{\sst{(8)}}}
\def\sst#1{{\scriptscriptstyle #1}}

\def\im{{\rm i}}

\begin{document}

\begin{flushright}
\hfill{KIAS-P12003}
\end{flushright}

\vspace{5pt}
\begin{center}
{\large {\bf $f(R)$ Theories of Supergravities and
Pseudo-supergravities}}

\vspace{10pt}

Haishan Liu\hoch{1}, H. L\"u\hoch{2,3} and Zhao-Long Wang\hoch{4}

\vspace{10pt}

\hoch{1}{\it Zheijiang Institute of Modern Physics\\
Department of Physics, Zhejiang University, Hangzhou 310027, China}

\vspace{10pt}

\hoch{2}{\it China Economics and Management Academy\\
Central University of Finance and Economics, Beijing 100081, China}

\vspace{10pt}

\hoch{3}{\it Institute for Advanced Study, Shenzhen University\\
Nanhai Ave 3688, Shenzhen 518060, China}

\vspace{10pt}

\hoch{4} {\it School of Physics, Korea Institute for Advanced Study,
Seoul 130-722, Korea}

\vspace{30pt}

\underline{ABSTRACT}
\end{center}

We present $f(R)$ theories of ten-dimensional supergravities,
including the fermionic sector up to the quadratic order in fermion
fields. They are obtained by performing the conformal scaling on the
usual supergravities to the $f(R)$ frame in which the dilaton
becomes an auxiliary field and can be integrated out.  The $f(R)$
frame coincides with that of M-theory, D2-branes or NS-NS 5-branes.
We study various BPS $p$-brane solutions and their near-horizon
AdS$\times$sphere geometries in the context of the $f(R)$ theories.
We find that new solutions emerge with global structures that do not
exist in the corresponding solutions of the original supergravity
description. In lower dimensions, We construct the $f(R)$ theory of
${\cal N}=2$, $D=5$ gauged supergravity with a vector multiplet, and
that for the four-dimensional $U(1)^4$ gauged theory with three
vector fields set equal. We find that some previously-known BPS
singular ``superstars'' become wormholes in the $f(R)$ theories. We
also construct a large class of $f(R)$ (gauged)
pseudo-supergravities.  In addition we show that the breathing mode
in the Kaluza-Klein reduction of Gauss-Bonnet gravity on $S^1$ is an
auxiliary field and can be integrated out.

\vspace{30pt} {\footnotesize{Emails: hsliu.zju@gmail.com\ \ \
mrhonglu@gmail.com\ \ \ zlwang4@gmail.com}}

\vspace{15pt}

\thispagestyle{empty}

\pagebreak

\tableofcontents

\addtocontents{toc}{\protect\setcounter{tocdepth}{2}}


\newpage

\section{Introduction}

     Under the principle of general coordinate transformation
invariance, there are limited ways of generalizing Einstein gravity
at the field theoretical level, by adding additional fields and/or
including higher-order curvature contributions. Largely motivated by
string theory, there have been decades of efforts in constructing
supergravities, which are believed to be the low energy effective
theories of string or more fundamental M-theory. The study of
extended gravities with higher-order curvature terms predated
supergravity and string theories.  The primary motivation was to
render the theory renormalizable by adding higher-order propagators.
It turns out that although gravity with higher-order curvature terms
can indeed be renormalizable, it suffers from having ghost degrees
of freedom \cite{stelle1,stelle2}. Recently there has been progress
in eliminating those ghosts at the classical level for some special
regions of the parameter spaces
\cite{strom,nmg,lpcritical,lppextension}; however, it is unlikely
that these parameter regions can survive the renormalization group
flow.

    If one is to consider only classical or semi-classical
generalizations of Einstein gravity, the simplest example is perhaps
the $f(R)$ theory in which the Ricci scalar in the Einstein-Hilbert
action is replaced by certain appropriate function $f$ of $R$.
Interestingly, there has been almost no overlap in the research
areas of supergravity and $f(R)$ gravity.  The work of $f(R)$
theories has been primarily focused on cosmology. (See reviews
\cite{rev0,rev1,rev2}.) Recently, we have demonstrated that there
exist a subclass of $f(R)$ theories that can admit Killing spinor
equations, which allow one to construct ``BPS'' solutions
\cite{llwfr}.  Whilst Killing spinor equations can exist in some
intrinsically non-supersymmetric theories \cite{lpw,lw,llw}, their
appearance is certainly an important characteristic of
supergravities. It was shown in \cite{lpwpseudo,llwpseudo} that when
a bosonic gravity theory admits Killing spinor equations, it can be
promoted to pseudo-supergravity which is invariant under the
pseudo-supersymmetric transformation rules up to the quadratic order
in fermion fields. This suggests that there should exist $f(R)$
pseudo-supergravities. Furthermore, if we add appropriate additional
fields so that the the degrees of freedom of the bosons and fermions
match, we may expect to obtain an $f(R)$ theory of supergravity.

     It is well known that $f(R)$ gravity is equivalent to a special
class of the Brans-Dicke theory in which the scalar field has no
kinetic term.  Conversely, any gravity theory coupled to a scalar
can be cast into the ``$f(R)$ frame'' in which the scalar has no
kinetic term and may become auxiliary.  Integrating out this
auxiliary scalar gives rise to the $f(R)$ theory.  Thus, the
relation between the $f(R)$ theory and the corresponding Brans-Dicke
theory is analogous to that between the Nambu-Goto and Polyakov
actions in the string theory.

In this paper, we study supergravities and perform the conformal
transformation so that the theories are in the $f(R)$ frame.  In
general the resulting equation of motion of the scalar field is a
polynomial of the scalar with irrational power. However, we find
that for all supergravities in $D=10$, the polynomial is of integer
power and hence the scalar can be straightforwardly integrated out,
giving rise to the $f(R)$ theory description of these
supergravities.  We find that this can also be done for ${\cal
N}=2$, $D=5$ gauged supergravity with a vector multiplet, and $D=4$
$U(1)^4$ gauged theory with three $U(1)$ vectors set to equal. It
should be emphasized that although the $f(R)$ theory is equivalent
to the Brans-Dicke description at the classical level, they are
inequivalent to the original supergravities even at the classical
level, since the conformal scaling can be singular in the solution
space. We obtain new solutions that are well behaved in the $f(R)$
theories but would be discarded in the original supergravities owing
to the bad properties.

The paper is organized as follows.  In section 2, we review $f(R)$
gravity and its connection to the special class of Brans-Dicke
theory where the scalar has no kinetic term.  We then study the
conversion of the gravity/scalar system to the $f(R)$ theory.  This
can be done by first performing a conformal scaling on the theory
from the Einstein frame to the $f(R)$ frame and then integrating out
the scalar. We present two examples of scalar potentials that appear
frequently in supergravities and obtain their corresponding $f(R)$
theories. In particular, we demonstrate that AdS worm-branes, domain
walls that connect two AdS Minkowski boundaries, can emerge in the
$f(R)$ theories.

    In section 3, we study the nature of the $f(R)$ frame. It turns
out that if the $D$-dimensional theory comes from the Kaluza-Klein
$S^1$ reduction of certain $(D+1)$-dimensional theory, the $f(R)$
frame is in fact the frame of the $(D+1)$-dimensional metric without
any conformal scaling.  We then consider a general Lagrangian in
arbitrary dimensions involving two vectors and a scalar with a
non-trivial scalar potential.  This Lagrangian reduces to special
cases of the $U(1)^3$ and $U(1)^4$ theories in the $D=5$ and $D=4$
gauged supergravities respectively.  We construct the $f(R)$ theory
of this system, and demonstrate that some singular ``black holes''
of the original theory become smooth wormholes in the $f(R)$ theory.

    In section 4, we construct $f(R)$ theories of ten-dimensional
supergravities, focusing on the bosonic sector.  We first cast
supergravities in the $f(R)$ frame in which the dilaton, which
measure the string coupling, becomes an auxiliary field. Integrating
out this scalar leads to the the $f(R)$ theories of supergravities.
For those ten-dimensional supergravities that comes from  M-theory
on $S^1$ or $S^1/\mathbb{Z}_2$ reductions, the $f(R)$ frame is
nothing but the M-theory frame, which also coincides with the
D2-brane frame and the NS-NS 5-brane frame.  We examine the
previously-known $p$-branes in the context of $f(R)$ theories and
show that some previously singular solutions are much better behaved
in the $f(R)$ description. For example the usual singular NS-NS
string now interpolates between the AdS$_3\times S^7$ horizon to the
asymptotic flat spacetime.  We also obtain new class of $p$-branes
which are supported by delta-function source located at the equator
of the foliating sphere in the transverse space.  We show that some
coordinate of such a solution in the $f(R)$ description has extended
range of that in the corresponding local solution in M-theory. This
suggests that new non-perturbative physical degrees of freedom can
be uncovered by the $f(R)$ theories.

     In section 5, we study the fermionic sector of $f(R)$
supergravities. We show that the dilaton remains an auxiliary field
even when the fermion sector is included. For ${\cal N}=1$, $D=10$
and type IIA supergravities, we give the fermionic Lagrangian and
the supersymmetric transformation rules up to the quadratic order in
fermion fields.  We also give the general structure of the $f(R)$
theory involving the gravitino and dilatino fields in general
dimensions.

      In section 6, we consider ${\cal N}=2$, $D=5$ gauged
supergravity with a vector multiplet.  We construct the
corresponding $f(R)$ gauged supergravity.  We obtain AdS worm-branes
and charged wormholes in the $f(R)$ gauged supergravity.  We extend
the discussion to $D=4$ gauged supergravity and also the gauged
$f(R)$ Kaluza-Klein pseudo-supergravity.  In section 7, we construct
a large class of $f(R)$ pseudo-supergravities in general dimensions.
We give the conclusions and present further discussions in section
8. In appendix A, we study the Kaluza-Klein circle reduction where
the lower-dimensional metric is not scaled by the breathing mode. We
demonstrate that the breathing mode is auxiliary even when the
higher-order Gauss-Bonnet curvature term is included.  In appendix
B, we present a general class of charged black hole solutions of the
theory discussed in section 3.2.

\section{Converting the gravity/scalar system to $f(R)$}

In this section, we shall give a quick review of $f(R)$ gravity,
which is defined by replacing the Ricci scalar $R$ in the
Einstein-Hilbert action with an appropriate function $f$ of $R$.
For non-vanishing $f''(R)$, the theory was known to be related to a
special class of the Brans-Dicke theory. We shall study this
relation focusing on the conversion of a gravity/scalar system to
$f(R)$ gravity. We consider two explicit examples of scalar
potentials that arise in supergravities. We demonstrate that the two
systems are not equivalent, owing to the possibility that singular
conformal scaling may arise in the solution space. In other words,
the $f(R)$ theories can have solutions with global properties that
do not exist in the corresponding local solutions of the
gravity/scalar systems of supergravities.

\subsection{Equations of motion}

The Lagrangian of $f(R)$ gravity in general dimensions is given by
\begin{equation}
e^{-1} {\cal L}_D=f(R)\,,
\end{equation}
where $e=\sqrt{-\det(g_{\mu\nu})}$. (There should be no confusion
between this $e$ which appears only as $e^{-1}$ in this paper and
the notation for the exponential function.) In this paper, we shall
be concerned with only $f(R)$ theories in the metric formalism, and
hence the equations of motion from the variation of $g_{\mu\nu}$ are
given by
\begin{equation}
{\cal G}_{\mu\nu}\equiv F(R) R_{\mu\nu} - \ft12 f(R) g_{\mu\nu} +
(g_{\mu\nu}\Box - \nabla_\mu\nabla_\nu) F(R)=0\,,\label{Gmunueom}
\end{equation}
where $F(R)=f'(R)$.  Note that in this paper, we always use a prime
to denote a derivative with respect to $R$, unless an explicit new
variable is given. Taking the trace, we have
\begin{equation}
{\cal R}\equiv RF - \ft12 D f + (D-1) \Box F=0\,.\label{traceeom}
\end{equation}
The equations of motion (\ref{Gmunueom}) can be equivalently
expressed as
\begin{equation}
{\cal R}_{\mu\nu}\equiv R_{\mu\nu} - \fft{1}{F} \nabla_\mu\nabla_\nu
F+ \fft{1}{2(D-1)F} ( f-2R F)g_{\mu\nu}=0\,.\label{Rmunueom}
\end{equation}

The simplest class of solutions for $f(R)$ gravities are perhaps the
Einstein metrics, for which $R$ is a constant, which we denote as
$R_0\equiv D\Lambda$, where $\Lambda$ is the effective cosmological
constant.  It follows from (\ref{traceeom}) that
\begin{equation}
2R_0 F(R_0)=Df(R_0)\,.\label{R0eom}
\end{equation}
Depending on whether $\Lambda$ is positive, 0, or negative, the
vacuum solution is de Sitter (dS), Minkowski and anti-de Sitter
(AdS) respectively.  In the special case when $f(R_0)=0=F(R_0)$, the
full equations of motion are reduced to simply the scalar type of
equation $R=R_0$, and the theory has no propagating spin-2 degrees
of freedom \cite{llwfr}.  It was shown in \cite{llwfr} that there
exists a subclass of $f(R)$ theories that admit Killing spinor
equations. Exact non-trivial ``BPS'' domain walls and cosmological
solutions with varying $R$ are consequently obtained. It was
demonstrated that $f(R)$ theories can admit Einstein metrics that do
not satisfy (\ref{traceeom}); they are characterized by the
divergent $F(R_0)$ \cite{llwfr}.

\subsection{The conversion}

It is well-known that $f(R)$ gravity can be cast into the form of
the Brans-Dicke theory. To see this, one starts with the Lagrangian
\begin{equation}
e^{-1}{\cal L}=f(\chi) + f_{,\chi}(\chi)
(R-\chi)\,.\label{legrender}
\end{equation}
Variation with respect to $\chi$ gives rise to
\begin{equation}
f_{,\chi\chi}(R-\chi)=0\,.
\end{equation}
Thus provided that $f_{,\chi\chi}\ne 0$, one has $\chi=R$, and hence
(\ref{legrender}) gives rise to the usual $f(R)$ theory.  On the
other hand, we can treat $\chi$ as a scalar field and hence the
Lagrangian (\ref{legrender}) is a gravity/scalar system.  To make
this manifest, one can define
\begin{equation}
\varphi=f_{,\chi}(\chi)\,,\label{converstion}
\end{equation}
and hence the $f(R)$ gravity is equivalent to the Brans-Dicke theory
of the type
\begin{equation}
e^{-1}{\cal L}=\varphi R + f(\chi(\varphi))- \varphi\,
\chi(\varphi)\,.\label{BDT}
\end{equation}
This is a special class of Brans-Dicke theory with no manifest
kinetic term for $\varphi$.  The conversion of $f(R)$ gravity to the
Brans-Dicke theory requires finding the inverse function of $F=f'$,
which in general does not have explicit analytical form.  The
absence of the kinetic term for $\varphi$ implies that the the
variation of $\varphi$ gives rise to a pure algebraic equation for
$\varphi$.

    Conversely, one may convert some gravity/scalar systems to $f(R)$
theories. In supergravities, the Lagrangians are typically written
in the Einstein frame.  For now we consider gravity coupled to a
single scalar only:
\begin{equation}
e^{-1}{\cal L}_D=R - \ft12(\partial\phi)^2 -
V(\phi)\,.\label{genscalarlag}
\end{equation}
Let us make the following conformal transformation and the field
redefinition for the dilaton
\begin{equation}
g_{\mu\nu} \rightarrow e^{-2\kappa \alpha\phi} g_{\mu\nu}\,,\qquad
\varphi = e^{\kappa \beta \phi}\,,\label{conformscaling}
\end{equation}
with $\kappa=\pm 1$ and
\begin{equation}
\alpha = -\fft{1}{\sqrt{2(D-1)(D-2)}}\,,\qquad \beta=
\sqrt{\fft{D-2}{2(D-1)}}\,.\label{alphabeta}
\end{equation}
It is clear that this conformal transformation can be done in all
dimensions greater or equal to three. Note that these two constants
are rational numbers only in $D=3$ and $D=10$. The Lagrangian
(\ref{genscalarlag}) becomes
\begin{equation}
e^{-1} {\cal L}_D=\varphi R - \widetilde
V(\varphi)\,,\label{widetildeV}
\end{equation}
where
\begin{equation}
\widetilde V(\varphi)=\varphi^{\fft{D}{D-2}} V(\phi(\varphi))\,,
\end{equation}
Comparing to (\ref{BDT}), we find that if the theory can be
converted into an $f(R)$ theory, $f$ must satisfy the following
\begin{equation}
f(R)- R F(R) + \widetilde V(F)=0\,.\label{convert}
\end{equation}
For the situation with non-vanishing $f''(R)$, acting with
$\partial_R$, we have
\begin{equation}
\fft{d\widetilde V}{dF} = R\,,\label{solvefortildeV}
\end{equation}
This is a purely algebraic equation for the function $F(R)$. Once we
solve for $F$, the $f(R)$ theories can be derived by a first-order
integration.

A more direct approach is to view the $\varphi$ in
(\ref{widetildeV}) as an auxiliary field since it does not have a
kinetic term.  The equation of motion associated with the variation
of $\varphi$ is $d\widetilde V/d\varphi=R$, which is an algebraic
equation for $\varphi$. Solving for $\varphi$ and substituting it
back into (\ref{widetildeV}), we obtain the corresponding $f(R)$
theory. Thus the relation between the $f(R)$ theory and the
Brans-Dicke theory (\ref{widetildeV}) is analogous to that between
the Nambu-Goto and Polyakov actions of the string theory, and hence
they are classically equivalent.  Of course, it is not always
possible to get a close form solution even for an algebraic
equation. However, it should be emphasized that the $f(R)$ theory
can be inequivalent to the original gravity/scalar system
(\ref{genscalarlag}) since the conformal scaling can be singular.
Note that $\varphi=F(R)$ and hence a necessary ghost-free condition
is that $F(R)$ is non-negative.

A special case should be addressed in which the potential function
$\widetilde V$ is linear in $\varphi$, {\it i.e.}
\begin{equation}
\widetilde V= a \varphi + b\,,\qquad \longrightarrow\qquad e^{-1}
{\cal L}_D= \varphi(R - a) - b\,.\label{nofrlag}
\end{equation}
Substituting $\widetilde V$ in (\ref{convert}), we have an $f(R)$
that is linear on $R$, {\it i.e.} the usual Einstein gravity with a
cosmological constant.  The converted theory is inequivalent to the
original theory, owing to the fact $f''(R)=0$ in this case. In fact
the equation of motion for $\varphi$ gives no information on
$\varphi$, but simply tells us that $R=a$.  Thus the gravity/scalar
system cannot be converted into $f(R)$ theory with $\varphi$
absorbed as $F$. However, in many explicit examples, the Lagrangian
(\ref{nofrlag}) can be viewed as a limiting one of a more general
$f(R)$ theory. The $f(R)$ limit to Einstein gravity with
$f''(R)\rightarrow 0$ was studied in \cite{olmo}.

It is worth remarking that in higher-derivative theories, we can
expect that $\widetilde V$ in (\ref{widetildeV}) is a function not
only of $\varphi$, but also of spacetime derivatives of $\varphi$.
Then, $\varphi$ ceases to be auxiliary.  Thus it is much more
non-trivial for $\varphi$ to be auxiliary in higher-derivative
gravities. Further discussion and an explicit example are provided
in appendix A.

As we have mentioned, for a generic function $V$, there is no
analytical solution to the equation (\ref{solvefortildeV}).  We
shall give two examples of $V$ that are relevant to this paper, for
which analytical solutions for $f$ can be found.

\subsection{Two examples}

In this subsection, we consider two classes of scalar potentials
that frequently appear in supergravities.

\bigskip
\noindent{\bf Example 1}:  The first example is just a pure
exponential potential
\begin{equation}
V(\phi) = \ft12 m^2 e^{-D\alpha\, b\,\phi}\,,\label{gensepot}
\end{equation}
where $\alpha$ is given in (\ref{alphabeta}) and $b$ is an arbitrary
constant parameter. The scalar potential in massive type IIA
supergravity \cite{massive2a} is of this form.  Some scalar
potentials in gauged supergravities also take this form but with
negative $m^2$. (See {\it e.g.},
\cite{fresch,salsez,chasab,cglpred}.) Furthermore, the dilaton
coupling with the form-fields is of this form if we can treat the
form-fields as constant. It is straightforward to obtain the
corresponding $f(R)$:
\begin{equation}
f(R)=\fft{2+\kappa D b}{D(1+\kappa b)} R\, \Big(\fft{2(D-2)R}{D(1
+\kappa b)m^2}\Big)^{\fft{D-2}{2 + \kappa D b}}\,.\label{frxxx}
\end{equation}
Note that there is an identity between $f$ and $F$, given by
\begin{equation}
f=\fft{2 + \kappa D b}{D(1 + \kappa b)}\,F\, R\,,\label{fFlag0}
\end{equation}
This identity allows us to take a smooth limit of sending
$m\rightarrow 0$ and recover the Einstein theory with a free scalar.
To see this, note that if we let $m$ vanish, the regularity of the
$f(R)$ theory (\ref{frxxx}) requires that $R$ approaches zero. There
is no unique way of how these two quantities approach zero, and the
ratio can be described by a scalar quantity.  If we define
$\varphi=F$, and in the $m\rightarrow 0$, it follows from
(\ref{fFlag0}) that the Lagrangian becomes
\begin{equation}
e^{-1} {\cal L} \sim \varphi R\,.
\end{equation}
This Lagrangian is consistent with the limit $R\rightarrow 0$ since
the variation of $\varphi$ gives rise to $R=0$.  Converting this
Lagrangian to Einstein frame, we obtain the original Einstein theory
with a free scalar.

In general the parameter $b$ in (\ref{gensepot}) is irrational in
supergravities.  The resulting $f(R)$ theory will have irrational
power of $R$.  We consider this as an unnatural formalism of a
gravity theory.  We shall focus our attention in examples where $b$
is rational. An interesting case is $b=0$.  The scalar/gravity
system is simply the cosmological Einstein gravity with a
``massless'' free scalar. The resulting $f(R)$ theory becomes
\begin{equation}
f(R)\propto R^{\fft{D}{2}}\,.
\end{equation}
This $f(R)$ is interesting in that all Einstein metrics with any
effective cosmological constants are solutions.

     It should be remarked that if $(D-2)/(2+\kappa Db)>0$, the
$f(R)$ theory (\ref{frxxx}) admits a solution $R=0$. Such a solution
clearly does not exist in the original gravity/scalar theory.

\bigskip
\noindent{\bf Example 2}:  The second example is a scalar potential
involving two specific exponential terms
\begin{eqnarray}
V&=&-(D-1)\Big((D-3) g_1^2 e^{2\alpha \phi} + g_2^2 e^{-2(D-3)
\alpha \phi}\Big) \cr 
&=& -(D-1)\Big( (D-3) g_1^2 e^{-\sqrt{\fft{2}{(D-1)(D-2)}}\,\phi} +
g_2^2 e^{\fft{\sqrt2\,(D-3)}{\sqrt{(D-1)(D-2)}}\,\phi}\Big)\,.
\label{f2pot}
\end{eqnarray}
One reason for us to consider this scalar potential is that as we
shall see in the next section, it can be embedded in various gauged
supergravities.  This scalar potential can be expressed in terms of
a superpotential $W$, namely \cite{llw}
\begin{eqnarray}
V&=&\Big(\fft{dW}{d\phi}\Big)^2 - \fft{D-1}{2(D-2)} W^2\,,\cr 
W&=&\fft{1}{\sqrt2} \Big( g^2_1g^{-1}_2(D-3)
e^{-\fft{D-1}{\sqrt{2(D-1)(D-2)}}\phi} + (D-1)g_2
e^{\fft{D-3}{\sqrt{2(D-1)(D-2)}}\phi}\Big)\,.
\end{eqnarray}
Under the conformal transformation (\ref{conformscaling}), the
Lagrangian becomes
\begin{equation}
e^{-1} {\cal L} = \varphi \Big(R + (D-1)(D-3)g_1^2\Big) +
(D-1)g_2^2\varphi^3\,.\label{gslag4}
\end{equation}
Note that if we simply set $g_2=0$, the theory cannot be converted
to the $f(R)$ formalism.  In general, it follows from
(\ref{solvefortildeV}), we have
\begin{equation}
F=\fft{\sqrt{-R -
(D-1)(D-3)g_1^2}}{\sqrt{3(D-1)g_2^2}}\,.\label{ntfr0}
\end{equation}
The corresponding $f(R)$ is given by
\begin{equation}
f=\ft23 F(R)\, \Big(R + (D-1)(D-3)g_1^2\Big)\,.\label{ntfr1}
\end{equation}
Thus we have obtained the $f(R)$ theory from the gravity/scalar
system (\ref{f2pot}).  In the limit of $g_2\rightarrow 0$, we
reproduce the Lagrangian (\ref{gslag4}) up to an overall scaling
factor $2/3$.

    Although the $f(R)$ theory (\ref{ntfr1}) with (\ref{ntfr0}) is
obtained from the gravity/scalar system (\ref{genscalarlag}) with
the potential (\ref{f2pot}), the two theories should not be
considered as equivalent, even at the classical level.  It follows
from (\ref{ntfr0}) and (\ref{ntfr1}) that we have $f(R_0)=0=F(R_0)$,
where $R_0=-(D-1)(D-3)g_1^2$.  Thus any metrics with $R=R_0$ is a
solution. Such a solution does not exist in the original
gravity/scalar system, demonstrating that the two theories are not
equivalent.

  As we shall discuss in section 6, it turns out that the $f(R)$
theory (\ref{ntfr1}) admits the following Killing spinor equations
\begin{equation}
D_\mu \epsilon+ \ft12 g F \Gamma_\mu \epsilon=0\,,\qquad \Gamma^\mu
\partial_\mu F\epsilon - \ft12 (D-3) (F^2-1)
\epsilon=0\,,\label{dwks}
\end{equation}
where we have set $g_1=g=g_2$.  Following the technique of
\cite{llwfr}, we find the following ``BPS'' domain-wall solution
\begin{equation}
ds^2 = dr^2 + \Big(\cosh(\ft12(D-3)g\, r)\Big)^{\fft4{D-3}}\, dx^\mu
dx_\mu\,,\qquad F=\tanh \Big(\ft12(D-3) g\,
|r|\Big)\,.\label{wormbrane}
\end{equation}
This solution describes an AdS worm-brane, connecting two AdS
Minkowski boundaries at $r=\pm \infty$ without a horizon in between.
Note the absolute-value sign on $r$ is imposed in $F$ to ensure the
solution is ghost free. It follows that a matter delta-function
source at $r=0$ is necessary to sustain this worm-brane.  In the
original gravity/scalar system, this AdS worm-brane would become a
singular domain wall with a power-law curvature singularity at
$r=0$.  Thus, although the local solutions are related by the
conformal transformation, the globally-defined worm-brane with $r$
running from $-\infty$ to $+\infty$ does not exist in the original
gravity/scalar theory.

\section{$f(R)$ theories of (gauged) Kaluza-Klein gravities}

In the previous section, we present the formalism and examples of
converting a system of gravity coupled with a single scalar to
$f(R)$ theories.  The necessary conformal scaling
(\ref{conformscaling}) is universal, independent of the scalar
potential, and the constants $(\alpha, \beta)$ are remarkably the
same as those associated with the scaling factors in the
Kuluza-Klein circle reduction \cite{stainless}.  It is thus of
interest to investigate the connection between the $f(R)$ frame and
the dimensional reduction.  Also we would like to investigate
whether we can convert some supergravity theories involving form
fields into the $f(R)$ formalism.  We shall study both questions by
examining the Kaluza-Klein theory that is $S^1$ reduction of pure
Einstein gravity. Starting from Einstein gravity in $(D+1)$
dimensions
\begin{equation}
\hat e^{-1}{\cal L}_{D+1} = \hat R\,,
\end{equation}
we perform the circle reduction with the metric ansatz
\cite{stainless}
\begin{equation}
d\hat s^2_{D+1} = e^{2\alpha\phi} ds_{\rm Ein}^2 + e^{2\beta\phi}
(dz+ {\cal A}_\1)^2\,,\label{reduction0}
\end{equation}
where $\alpha$ and $\beta$ are given by (\ref{alphabeta}). The
metric $ds_{\rm Ein}^2$ in lower dimensions is in the Einstein
frame. The resulting $D$-dimensional Lagrangian is
\begin{equation}
e^{-1} {\cal L} = R - \ft12(\partial \phi)^2 - \ft14
e^{-2(D-1)\alpha\phi} {\cal F}_\2^2\,.\label{KKlag1}
\end{equation}
where ${\cal F}_\2=d{\cal A}_\1$.  What we like to draw attention
here is that the conformal transformation (\ref{conformscaling})
implies that
\begin{equation}
d\hat s^2_{D+1} = e^{2(1-\kappa)\alpha\phi} ds_D^2 + e^{2\beta\phi}
(dz+ {\cal A}_\1)^2\,.
\end{equation}
For $\kappa=+1$, we have
\begin{equation}
d\hat s^2_{D+1} = d\hat s_D^2 + \varphi^2 (dz+ {\cal
A}_\1)^2\,.\label{reduction1}
\end{equation}
Thus the metric in $D$ dimensions is in the same frame as that in
$(D+1)$ dimensions, without any conformal scaling.  For $\kappa=-1$,
The reduction ansatz becomes
\begin{equation}
ds_{D+1}^2 = \varphi^{\fft4{D-2}} ds_D^2+ \varphi^{-2} (dz +{\cal
A}_\1)^2\,,\label{reduction2}
\end{equation}
This does not appear to have a particular interesting physical
interpretation, and hence we shall focus our attention primarily on
$\kappa=+1$.  Thus if a $D$-dimensional theory has an origin in
$(D+1)$ dimensions, we may define the ``$f(R)$ frame'' simply as the
$(D+1)$-dimensional frame.

\subsection{$f(R)$ Kaluza-Klein gravity}

With the reduction ansatz (\ref{reduction1}), the Lagrangian of
Kaluza-Klein gravity in $D$ dimensions in the $f(R)$-frame is
\begin{equation}
e^{-1}{\cal L}_D= \varphi R - \ft14 \varphi^3 {\cal
F}_\2^2\,.\label{KKlag2}
\end{equation}
Thus we see that the breathing mode $\varphi$ is an auxiliary scalar
field. In appendix A, we present the Kaluza-Klein reduction with the
metric ansatz (\ref{reduction1}).  We show that when a generic
higher-order curvature term is included in $(D+1)$ dimensions, the
scalar $\varphi$ ceases to be auxiliary. However, it remains an
auxiliary field for the circle reduction of Gauss-Bonnet gravity.

In the procedure of converting the Kaluza-Klein theory
(\ref{KKlag2}) to an $f(R)$ theory, we can treat ${\cal F}_\2^2$ as
if it is a constant. The Lagrangian is then analogous to the first
example discussed in section 2.3. It follows from the discussion
there that $F(R)=2 \sqrt{R/(3{\cal F}_\2^2)}$. Thus the $f(R)$
theory of the Kaluza-Klein gravity (\ref{KKlag1}) is given by
\begin{equation}
e^{-1}{\cal L}_D=f(R)=\fft43\sqrt{\fft{R^3}{3{\cal
F}_\2^2}}\,.\label{kkfrlag}
\end{equation}
To demonstrate that this derivation with ${\cal F}_\2^2$ being
treated as a constant is legitimate, we give the two equations of
motion associated with $\delta {\cal A}_\mu$ and $\delta
g^{\mu\nu}$:
\begin{eqnarray}
&&\nabla_\mu (F^3 {\cal F}^{\mu\nu}) =0\,,\cr 
&&F R_{\mu\nu} - \ft12 f g_{\mu\nu} + (g_{\mu\nu} \Box -
\nabla_\mu\nabla_\nu) F - \ft12 F^3\, ({\cal F}^2_\2)_{\mu\nu}=0\,.
\label{KKeom1}
\end{eqnarray}
It is easy to verify that these equations of motion are the same as
the ones derive from (\ref{KKlag2}), and hence they are also
equivalent to those from (\ref{KKlag1}), up to the conformal scaling
which may be singular.  Note that in this paper, we use notation $f$
and $F$ exclusively for $f(R)$ and $F(R)$.  The notation $F$ should
not be confused with form fields which either carry explicit indices
or the subscript indicating the rank of the form.

     It is worth remarking that to be pedantic the proper $f$
expression should be $f=4/3 R \sqrt{R/(3{\cal F}_\2^2)}$.  Since
$\sqrt{R^2}=\pm R$ depending on the sign of $R$, there may be an
overall minus sign in the Lagrangian (\ref{kkfrlag}).  In this
paper, we shall not be always precise regarding this overall sign of
the Lagrangian, unless the issue of the ghost-free condition is
discussed.

If the $(D+1)$-dimensional Einstein gravity is coupled to a
cosmological constant, namely
\begin{equation}
\hat e^{-1} {\cal L}_{D+1} = \hat R - \Lambda_0\,,
\end{equation}
The reduced theory becomes
\begin{equation}
e^{-1}{\cal L}_D= \varphi (R-\Lambda_0) - \ft14 \varphi^3 {\cal
F}_\2^2 \,.\label{KKlag3}
\end{equation}
The corresponding $f(R)$ is given by
\begin{equation}
f(R)=\fft{4}{3} \sqrt{-\fft{(\Lambda_0 - R)^3}{{\cal F}_\2^2}}\,.
\end{equation}
The form of the equations of motion is identical to (\ref{KKeom1}).

    A natural question to ask is what happens when ${\cal F}_\2$
vanishes.  In this case, the regularity of the action requires that
$R=\Lambda_0$.  It is clear that there is no unique way how the two
quantities ${\cal F}_\2$ and $(R-\Lambda_0)$ vanish.  There should a
scalar field describing the ratio of these two quantities when they
approaches zero.  Considering the identity
\begin{equation}
f(R) = \ft23 F(R)\, (R-\Lambda_0)\,,
\end{equation}
for any ${\cal F}_\2$, it is clear that $F=\varphi$ is the scalar
field, and hence we recover (\ref{KKlag3}) with ${\cal F}_\2=0$.  In
this case, the constraint $R=\Lambda_0$ is the equation of motion
for the $\varphi$ field.

Finally, let us consider $(D+1)$-dimensional Einstein gravity
coupled to an $n$-form field strength
\begin{equation}
\hat e^{-1} {\cal L}_{D+1} = \hat R - \ft{1}{2\,n!} \hat
F_{\sst{(n)}}^2\,.
\end{equation}
It is straightforward to see that the Kaluza-Klein theory in the
$f(R)$ frame is given by
\begin{eqnarray}
e^{-1} {\cal L} &=& \varphi (R - \ft1{2\,n!} F_{\sst{(n)}}^2) -
\ft1{2\,(n-1)!} \varphi^{-1} F_{\sst{(n-1)}}^2 - \ft14 \varphi^3
{\cal F}_\2^2\,,\cr 
F_{\sst{(n)}}&=& dA_{\sst{(n-1)}} - dA_{\sst{(n-2)}}\wedge {\cal
A}_\1\,,\qquad F_{\sst{(n-1)}}=dA_{\sst{(n-2)}}\,,\qquad {\cal
F}_\2=d{\cal A}_\1\,.\label{nformred}
\end{eqnarray}
Applying this result to M-theory, it can be easily deduced that the
$f(R)$ frame for ten-dimensional supergravities is the same as that
of the M-theory, D2-branes or NS-NS 5-branes.

\subsection{$f(R)$ gauged KK gravity and charged wormholes}

It was demonstrated that the Kaluza-Klein theory in any dimensions
can be pseudo-supersymmetrized by the inclusion of a
pseudo-gravitino and dilatino. The full Lagrangian is invariant
under the pseudo-supersymmetric transformation rules up to the
quadratic fermion order \cite{llwpseudo}.  Furthermore, the
pseudo-supergravity can be gauged and the fermions are all charged
under the Kaluza-Klein vector. The gauging generates a scalar
potential (\ref{f2pot}). The full bosonic Lagrangian is given by
\cite{llw,llwpseudo}
\begin{equation}
e^{-1} {\cal L} = R - \ft12(\partial \phi)^2 - \ft14
e^{-2(D-1)\alpha\phi} {\cal F}_\2^2-(D-1) g^2 \Big((D-3) e^{2\alpha
\phi} + e^{-2(D-3) \alpha \phi}\Big) \,,\label{gaugedKKlag}
\end{equation}
where $\alpha$ is given in (\ref{alphabeta}). Note that the scalar
potential was discussed as the second example in section 2.3. This
theory can be embedded in gauged supergravities in $D=4,5$ and 7
\cite{llw}. (See also, for example, \cite{tenauthor}.) In the case
of $D=6$, it may also be possible to embed the theory in the $F(4)$
gauged supergravity \cite{romans} coupled to a vector multiplet
\cite{cglp,clpbubble}. Under the conformal scaling with $\kappa=+1$,
we have
\begin{equation}
e^{-1} {\cal L} = \varphi \Big (R + (D-1)(D-3) g^2\Big) + \varphi^3
\Big(-\ft14 {\cal F}_\2^2 + g^2(D-1)\Big)\,,
\end{equation}
The corresponding $f(R)$ is given by
\begin{equation}
f(R)=\fft{4}{3\sqrt3} \sqrt{\fft{\Big(-R - (D-1)(D-3)g^2\Big)^3}{
4(D-1) g^2 - {\cal F}_\2^2}}\,.\label{KKgaugefrlag1}
\end{equation}
Again, the form of the equations of motion is identical to
(\ref{KKeom1}).

In fact the Lagrangian (\ref{gaugedKKlag}) can be generalized to
include another vector and a tensor as well, giving
\begin{eqnarray}
e^{-1} {\cal L} &=& R - \ft12 (\partial\phi)^2 - \ft1{12}
e^{-4\alpha \phi} F_\3^2 -\ft14 e^{2(D-3)\alpha \phi} F_\2^2- \ft14
e^{-2(D-1)\alpha \phi} {\cal F}_\2^2 \cr 
&&+ (D-1)g^2\Big((D-3) e^{-\sqrt{\fft{2}{(D-1)(D-2)}} \,\phi}+
e^{\fft{\sqrt2\,(D-3)}{\sqrt{(D-1)(D-2)}}
\,\phi}\Big)\,,\label{gaugegenlag}
\end{eqnarray}
where $F_\3=dA_\2 -{\cal A}_\1 \wedge dA_\1$, $F_\2=dA_\1$ and
${\cal F}_\2=d{\cal A}_\1$.  If we set $g=0$, the theory is the
$S^1$ reduction of Einstein gravity coupled to a 3-form field
strength in $(D+1)$ dimensions, whose corresponding $f(R)$ theory
was given in (\ref{nformred}) with $n=3$.  One reason that we are
interested in this theory is that it admits multi-charge AdS black
hole solutions, as we shall demonstrate in appendix B.  Furthermore,
as we shall discuss in section 6, this Lagrangian can also be
embedded in both $D=4$ and $D=5$ gauged supergravities. In the
$f(R)$ frame, the theory is given by
\begin{equation}
e^{-1}{\cal L}= \varphi \Big (R + (D-1)(D-3) g^2 - \ft1{12}
F_\3^2\Big) + \varphi^3 \Big(-\ft14 {\cal F}_\2^2 + g^2(D-1)\Big)
-\ft14 \varphi^{-1} F_\2^2\,. \label{gaugegenfr}
\end{equation}
The variation of $\varphi$ gives rise to a quadratic equation of
motion for $\varphi^2$, and hence the $f(R)$ theory can be obtained
straightforwardly. The expression is clumsy and we shall not present
it here.  If we set $F_\2=0$, the $f(R)$ takes the similar form as
(\ref{KKgaugefrlag1}) with the $F_\3^2$ term appropriately inserted.
If instead ${\cal F}_\2$ is such that ${\cal F}_\2^2=4g^2(D-1)$, we
have
\begin{eqnarray}
f(R) &=& F\, \Big(R + (D-1)(D-3) g^2- \ft1{12} F_\3^2\Big)\,,\cr 
F &=& \sqrt{-\fft{F_\2^2}{4\Big(R + (D-1)(D-2) g^2-\ft1{12}
F_\3^2\Big)}}\,.\label{KKgaugefrlag2}
\end{eqnarray}

As we have mentioned, in appendix B, we shall give a general class
of non-extremal static multi-charged black hole solutions for the
system (\ref{gaugegenfr}). The non-extremal parameter can be turned
off while keeping the charge parameters fixed.  The resulting
solution is given by
\begin{eqnarray}
ds_D^2&=&-{\cal H}^{-\fft{D-3}{D-2}} H^{-\fft{D-1}{D-2}}\, h\, dt^2
+ {\cal H}^{\fft{1}{D-2}} H^{\fft{D-1}{(D-2)(D-3)}}
\Big(\fft{dr^2}{h} + r^2 d\Omega_{D-2}^2\Big)\,,\cr 
{\cal F}_\2&=& dt\wedge d{\cal H}^{-1}\,,\qquad
F_\2=\sqrt{\fft{D-1}{D-3}}\, dt\wedge dH^{-1}\,,\cr 
h &=& 1 + g^2 r^2 {\cal H} H^{\fft{D-1}{D-3}}\,,\qquad e^{\phi} =
\Big(\fft{\cal H}{H}\Big)^{\sqrt{\fft{D-1}{2(D-2)}}}\,,\cr 
{\cal H} &=& 1 + \fft{\tilde q}{r^{D-3}}\,,\qquad H=1 +
\fft{q}{r^{D-3}}\,.\label{cbh1}
\end{eqnarray}
In the $f(R)$ frame, the metric becomes much simpler:
\begin{equation}
ds^2_D=-({\cal H} H)^{-1} h dt^2 + H^{\fft2{D-3}} (h^{-1} dr^2 + r^2
d\Omega_{D-2}^2)\,.\label{cbh2}
\end{equation}
The solution has a naked curvature power-law singularity in both
frames. This is because $r=0$ is not a horizon, but instead the
metric has the form $ds^2\sim r^{2(D-4)} dr^2 - dt^2 +
d\Omega_{D-2}^2$, and hence the natural radial coordinate is $\rho
=r^{D-3}$.  When $\rho$ becomes negative such that either $H$ or
${\cal H}$ vanishes, the metric has a power-law curvature
singularity.  As we shall discussed in section 6, in $D=4,5$, the
theory is part of gauged supergravities and these solutions are BPS
and called ``superstars.''

     Now if instead we set $\tilde q=0$, the metric in
the $f(R)$ frame (\ref{cbh2}) near $r=0$ behaves like $ds^2\sim
r^{D-5} dr^2 - dt^2 + d\Omega_{D-2}^2$ and thus the natural
coordinate should be $\rho=r^{(D-3)/2}$.  In terms of the $\rho$
coordinate the solution is given by
\begin{eqnarray}
ds_D^2 &=& -\fft{h}{\rho^2 + q^2} dt^2 + (\rho^2 +
q^2)^{\fft{2}{D-3}} \Big( \fft{4d\rho^2}{(D-3)^2 h} +
d\Omega_{D-2}^2\Big)\,,\cr 
A_\1&=&\sqrt{\fft{D-1}{D-3}}\,\fft{q}{\rho^2 + q} dt\,,\qquad
h=\rho^2 + g^2 (\rho^2 + q)^{\fft{D-1}{D-3}}\,,\qquad \varphi\equiv
F=\fft{|\rho|}{\rho^2 + q^2}\,,\label{chargwormhole}
\end{eqnarray}
It is clear that this solution describes a wormhole with the radial
coordinate $\rho$ running from $-\infty$ to $+\infty$, connecting
two $\mathbb{R}^t \times S^{D-2}$ boundaries.  The positivity
condition for $\varphi\equiv F$ require that an absolute-value sign
be added on $\rho$ in its expression, and hence the Einstein
equations of motion (\ref{Gmunueom}) requires that a delta-function
matter source be needed to support this wormhole. Note that the
level surfaces for this wormhole are $\mathbb{R}^t\times S^{D-2}$,
unlike the worm-brane discussed earlier.  If we convert this
solution to that of the original (\ref{gaugegenlag}) theory, then it
has a power-law curvature singularity at $\rho=0$.  Thus the charged
wormhole of our $f(R)$ theory with $r\in (-\infty,+\infty)$ does not
exist in the original theory (\ref{gaugegenlag}).

\section{$f(R)$ supergravities in $D=10$: the bosonic sector}

Having addressed the preliminaries in sections 2 and 3, we now turn
our attention to converting supergravities to the $f(R)$ formalism.
Eleven-dimensional supergravity \cite{11d} can be argued as the most
fundamental one; however, since there is no scalar mode in presence,
there can be no $f(R)$ formalism. In this section, we consider
$f(R)$ supergravities in ten dimensions. We shall focus our
attention only on the bosonic sector. The discussion of fermions
will be given in section \ref{fermionsection}.

\subsection{${\cal N}=1$, $D=10$ $f(R)$ supergravity}

The simplest supergravity in $D=10$ has ${\cal N}=1$ supersymmetry
\cite{bvn}. The field content for the bosonic sector consists of the
metric, a dilaton $\phi$ and a 2-form antisymmetric tensor field
$A_{\mu\nu}$. The bosonic Lagrangian is given by
\begin{equation}
e^{-1} {\cal L}_{10} = R - \ft12 (\partial \phi)^2 - \ft1{12}
e^{-\phi} F_\3^2\,,\label{n1d10lag1}
\end{equation}
Making a conformal transformation (\ref{conformscaling}) with
$\kappa=+1$, we have the theory in the $f(R)$ frame
\begin{equation} e^{-1} {\cal L}_{10} = \varphi\, R
- \ft1{12} \varphi^{-1} F_\3^2\,,\label{n1d10lag2}
\end{equation}
and hence the corresponding $f(R)$ theory is
\begin{equation}
e^{-1} {\cal L}_{10} = f(R)= \sqrt{-\ft13 R
F_\3^2}\,.\label{n1d10frlag}
\end{equation}
The equations of motion are given by
\begin{eqnarray}
&& F R_{\mu\nu} - \ft12 f g_{\mu\nu} + (g_{\mu\nu} \Box -
\nabla_\mu\nabla_\nu) F - \ft14 F^{-1}\, (F^2_\3)_{\mu\nu}=0\,, \cr
&&\nabla_{\mu} (F^{-1} F^{\mu\nu\rho}) =0\,.\label{n1d10freom}
\end{eqnarray}
Note that we have an identity $f=2F\, R$.  This implies that in the
limit of $F_\3^2$ goes to zero, the quantity $F$ should be viewed as
a scalar quantity that is held fixed, leading to (\ref{n1d10lag2})
with the vanishing 3-form.  The Hodge dual of $F_\3$ is a 7-form
$F_\7$.  The dual description of (\ref{n1d10lag2}) is given by
\begin{equation}
e^{-1}{\cal L}_{10} = \varphi (R - \ft1{10080}
F_\7^2)\,.\label{n1d10dual}
\end{equation}
This implies that the $f(R)$ frame is in fact that of NS-NS
5-branes.

Ten-dimensional ${\cal N}=1$ supergravity can be viewed as part of
the lower-energy effective theory of string.  The dilaton $\phi$
plays the role as the string loop expansion field.  The string
coupling constant is $g_s=<\!\!e^{\phi}\!\!>$.  Since $\varphi =
e^{\fft43\phi}$, it follows that we have
\begin{equation}
<\!\!\varphi\!\!> = g_s^{4/3}\,.
\end{equation}
An important feature in the theory (\ref{n1d10lag2}) is that the
scalar $\varphi$ is no longer a dynamical field of an independent
degree of freedom, but instead it is given algebraically by
$\varphi=\sqrt{-F_\3^2/(12R)}$.  As we mentioned in section 2, the
relation between the $f(R)$ theory (\ref{n1d10frlag}) and the
Brans-Dicke like theory (\ref{n1d10lag2}) is analogous to that of
the Nambu-Goto and Polyakov actions.  The description of
(\ref{n1d10frlag}) is of the non-perturbative nature with the scalar
associated with the loop expansions absorbed as part of $f(R)$.
Whilst the two theories (\ref{n1d10frlag}) and (\ref{n1d10lag2}) are
classically equivalent, they are not equivalent to the original
supergravity (\ref{n1d10lag1}) since the conformal scaling can be
singular.

It should be remarked that it is perhaps a misnomer to continually
call (\ref{n1d10frlag}) as an $f(R)$ theory, since it also contains
matter field $F_\3=dA_\2$, which {\it is not} a constant, but a
dynamical field. Nevertheless, we shall continue use the terminology
$f(R)$ owing to the lack of any elegant alternative.  Note that in
our $f(R)$ supergravities, the matter field $F_\3$ couples to
gravity through the scalar curvature rather than only {\it via} the
metric, which was typically considered in the literature.

     Having obtained the $f(R)$ theory of (\ref{n1d10lag1}), we
investigate the corresponding electric string and magnetic 5-brane
solutions.  Such $p$-brane solitons were extensively studied in
supergravities. (See {\it e.g.}~\cite{dklsoliton}.)  In addition to
reviewing these solutions in the $f(R)$ frame, we shall also
consider new $p$-brane solutions that would be discarded in the
usual supergravity discussions.

\bigskip
\noindent{\bf Eletric String:} Our $f(R)$ theory admits the
following electric string solution
\begin{eqnarray}
ds^2 &=& H^{-\fft23} (-dt^2 + dx^2) + H^{\fft13} dy^i dy^i\,,\cr 
F_\3 &=& dt\wedge dx \wedge dH^{-1}\,,\qquad \partial_i\partial_i
H=0\,.
\end{eqnarray}
The metric of this solution is simply a certain conformal scaling of
the NS-NS string solution.  The isotropic one is given by
\begin{equation}
dy^i dy^i = dr^2 + r^2 d\Omega_7^2\,,\qquad H=1 + q/r^6\,,
\end{equation}
Thus we have
\begin{equation}
F_\3^2 = 6 H^{-3} H'^2=\fft{6^3 q^2\,r^4}{(r^6+q)^3}\,\qquad
F=H^{-\fft13}\,.
\end{equation}
In $f(R)$ supergravity, the solution is regular from the horizon
$r=0$, which is AdS$_\3\times S^7$ to the asymptotic $r=\infty$ flat
spacetime. From the second equation (\ref{n1d10freom}), we can
define the conserved electric string charge
\begin{equation}
Q_1=\int_{r\rightarrow \infty} F^{-1} {*F_\3} = 7 q \omega_7\,,
\end{equation}
where $\omega_7$ is the volume of the unit $S^7$.  In this paper, we
denote $\omega_n$ and $\Omega_{\sst{(n)}}$ as the volume and the
volume form of the unit $S^n$ respectively. The ``1'' in the
harmonic function can be dropped, giving rise to the AdS$_3\times
S^7$ solution
\begin{eqnarray}
ds^2= \ell^2 \Big( \fft{dr^2}{r^2} + \ell^{-6} r^4 (-dt^2 + dx^2) +
d\Omega_\3^2\Big)\,,\qquad F_\3=6\ell^{-6} r^5 dt\wedge dx\wedge
dr\,.\label{ads3s7sol1}
\end{eqnarray}
It should be pointed out that this solution cannot be obtained by
the decoupling limit that is typically considered in the context of
the AdS/CFT correspondence.  If we scale $r\rightarrow \epsilon\, r$
and $x^\mu \rightarrow \epsilon^3 x^\mu$, and then send $\epsilon$
to zero, we find that although the metric becomes AdS$_3\times S^7$,
the quantity $F_\3$ blows up.  This is related to the fact that
quantities $f(R)$ and $F_\3^2$, $F(R)$ in (\ref{ads3s7sol1}) are all
divergent in the AdS boundary. However, this does not affect the
earlier statement that the string solution interpolates smoothly
between the AdS$_3\times S^7$ horizon and the asymptotic flat
spacetime.

      In the $F_\7$ dual description, where $F_\7$ is magnetic, the
metric of the solution is identical, but with $F_\7=6q \Omega_7$. In
this case, the AdS$_\3\times S^7$ can be obtained from the
decoupling limit, with both $F_\7^2$ and $f(R)$ finite, if one
overlooks the divergence of $F(R)$ in this limit.

       If we trace back to $D=11$, the string solution becomes the usual
M2-brane, and the AdS$_3\times S^7$ becomes AdS$_4\times S^7$. Since
half of the Killing spinors in AdS$_4\times S^7$ depend on the
world-volume coordinate \cite{lptads}, it follows that the
AdS$_3\times S^7$ preserves half of the supersymmetry, with no
supersymmtry enhancement.  Further discussion of supersymmetry will
be given in section 5.

\bigskip
\noindent{\bf Magnetic 5-brane}:  There are two types of magnetic
5-brane solutions. The first type is given by
\begin{eqnarray}
ds^2_{10} &=& H^{-\fft13} dx^\mu dx_\mu + H^{\fft23} (dr^2 + r^2
d\Omega_\3^2)\,,\cr 
F_\3 &=& 2q \Omega_\3\,,\qquad H=1 + \fft{q}{r^2}\,.
\end{eqnarray}
The solution carries the magnetic charge
\begin{equation}
Q_5=\int F_\3 = 4Q \pi^2\,.
\end{equation}
This solution is effective the usual NS-NS 5-brane written in the
new frame of the $f(R)$ theory.  Since $F_\3$ is constantly
proportional to $\Omega_\3$, the isometry of $S^3$ is preserved. The
solution however suffers a curvature power-law singularity at $r=0$.
Since the function $H$ is a harmonic function in the transverse
space, the solution can be generalized to a solution describing
multi-center 5-branes.  We can lift the solution back to $D=11$ and
obtain the smeared M5-brane.  Such $p$-brane solutions and their
behavior in different frames were discussed extensively in
\cite{dklsoliton}.

The theory in fact admits the second type of magnetic 5-brane
solutions that were not considered previously:
\begin{eqnarray}
ds_{10}^2 &=& H^{-\fft13} dx^\mu dx_\mu + H^{\fft23} (dr^2 + r^2
d\Omega_3^2)\,,\qquad d\Omega_3^2 = d\theta^2 + \sin^2\theta
d\Omega_2^2 \cr 
F_\3 &=& 3q \cos\theta\, \sin^2\theta\, d\theta\wedge \Omega_\2\,,
\qquad H=1 + \fft{q}{r^3}\,,
\end{eqnarray}
In contrast to the previous 5-brane solution, the metric in this
solution is smooth running from AdS$_7\times S^3$ at the $r=0$
horizon to the asymptotic flat spacetime at $r=\infty$. It should be
mentioned that in this solution $F_\3$ is not constant proportional
to $\Omega_\3$, the volume form of the $S^3$, and the function $H$
is not the harmonic function in the transverse space. Thus, the
solution does not generalize to the multi-center harmonic solutions.

      Unlike the previous string solution, there is a decoupling
limit for which the ``1'' in the function $H$ can be dropped.  This
can be done first by scaling $r\rightarrow \epsilon\, r$ and $x^\mu
\rightarrow \sqrt{\epsilon}\, x^\mu$ and then sending $\epsilon$ to
zero. The resulting AdS$_7\times S^3$ solution is given by
\begin{equation}
ds^2 = \ell^2 \Big(\fft{dr^2}{r^2} + r\, dx^\mu dx_\mu +
d\Omega_\3^2\Big)\,,\qquad F_\3=3\ell^3 \cos\theta\, \sin^2\theta
d\theta\wedge \Omega_\2\,,
\end{equation}

Note also that the solution has the following properties
\begin{equation}
F_\3=3q\cos\theta\, \Omega_\3\,,\qquad
F(R)=(r^3+q)^\fft13\cos\theta\,.
\end{equation}
The fact that $F_\3$ is not constantly proportional to the $S^3$
volume form implies that not all the $S^3$ isometry is preserved;
only the $SO(3)$ of $S^2$ is preserved. Also $F(R)$ vanishes at
$\theta=\pi/2$. Thus the solution would have a naked power-law
curvature singularity at $\theta=\pi/2$ if it were to be converted
to that of the usual ${\cal N}=1$, $D=10$ supergravity. From the
point of view of the ${\cal N}=1$, $D=10$ supergravity, the
coordinate $\theta$ can only run from $0$ to $\pi/2$ and hence the
5-brane charge is given by $Q_5=\int F_\3 = 4\pi q$.

   However, from the point of view in the $f(R)$ theory, the geodesic
completeness of the metric requires that the coordinate $\theta$
extend to include the region $[\pi/2,\pi]$ as well. This then
implies that that $Q_5=4\pi q - 4\pi q=0$. Furthermore, the function
$F(R)$ becomes negative in $\theta \in (\pi/2,\pi]$. Such a problem
can be averted by introducing a source such that
\begin{equation}
F_\3=3q|\cos\theta|\, \Omega_\3\,,\qquad F(R)=(r^3+q)^\fft13\,
|\cos\theta|\,.
\end{equation}
Having done that, the equation for the form field in
(\ref{n1d10freom}) is still exactly satisfied; however, the Einstein
equations produce a delta-function source at the equator
$\theta=\pi/2$.  Such a source on the equator is not uncommon in
supergravity solutions.  The embedding of the AdS$_6\times S^4$ of
the the localized D4/D8-brane \cite{youm} in massive type IIA
supergravity has a more serious power-law singularity at the equator
of the $S^4$. (See section 4.4.) With this set up, the magnetic
charge is doubled, namely
\begin{equation}
Q_5=\int F_\3 = 8\pi q\,.\label{nsnsfrq5}
\end{equation}
Thus we see that not only the global structures of the metrics in
$f(R)$ and original supergravities are different.  The magnetic
charges are different too and hence there is no reason to claim that
these two solutions are equivalent even though they are related
locally by the conformal scaling. Note that in the dual $F_\7$
description, the 5-brane carries the electric flux with
$F_\7=d^6x\wedge dH^{-1}$.

If we trace back the local solution to $D=11$, it becomes the usual
M5-brane with the metric
\begin{equation}
ds^2_{11} = ds_{10}^2 + F^2 d\psi^2=H^{-\fft13} dx^\mu dx_\mu +
H^{\fft23} (dr^2 + r^2 d\Omega_3^2 + r^2 \cos^2\theta\,d\psi^2)\,,
\end{equation}
where $\psi$, with period $2\pi$, is the internal coordinate. Since
$F$ appears in $D=11$ only as $F^2$, the absolute-value sign on $F$
drops. From the eleven-dimensional point of view,  the latitude
angle $\theta$ clearly runs from 0 to $\pi/2$. In the $f(R)$ theory,
however, we must extend the $\theta$ range to $[0,\pi]$ by
introducing a $\delta$-function source.  Thus, even if the local
solution of the 5-brane is the same as that from $D=11$, it
describes a different physical state from the M5-brane.  Thus
spectrum of the $f(R)$ theory contains states that may not
apparently exist in the narrow picture of M-theory.

Finally we would like to mention that if we take $\kappa=-1$ for the
conformal scaling (\ref{conformscaling}), the Lagrangian becomes
\begin{equation}
e^{-1} {\cal L}_{10} = \varphi\, R - \ft12 \varphi^2 F_\3^2\,.
\end{equation}
The corresponding $f(R)$ becomes
\begin{equation}
f(R)=\fft{3R^2}{F_\3^2}\,.\label{kappaminusfr}
\end{equation}
Interestingly, as we shall see later, the $f(R)$ theory associated
with the R-R 3-form field strength in type IIB supergravity takes
this form.

\subsection{$f(R)$ heterotic supergravity}

${\cal N}=1$, $D=10$ supergravities with additional $E_8\times E_8$
or $SO(32)$ Yang-Mills fields are the low-energy effective theory of
the corresponding heterotic string theories. The bosonic Lagrangian
for heterotic supergravity is given by
\begin{equation}
e^{-1} {\cal L}_{10} = R - \ft12 (\partial \phi)^2 - \ft1{12}
e^{-\phi} F_\3^2-\ft14 e^{-\fft12\phi} (F^I_\2)^2\,,
\end{equation}
where $F_\2^I$ are the field strengths for the Yang-Mills fields.
The 3-form field strength satisfies the Bianchi identity
\begin{equation}
dF_\3=\ft12 F_\2^I\wedge F_\2^I\,.
\end{equation}
The conformal transformation (\ref{conformscaling}) with $\kappa=+1$
leaves the Yang-Mills fields decoupled from the scalar, namely
\begin{equation}
e^{-1} {\cal L}_{10} = \varphi\, R - \ft12 \varphi^{-1} F_\3^2
-\ft14 (F^I_\2)^2\,,
\end{equation}
It follows that the bosonic Lagrangian of the $f(R)$ heterotic
supergravity is given by
\begin{equation}
e^{-1} {\cal L}_{10} = \sqrt{-\ft13 R F_\3^2} - \ft14 (F^I_\2)^2\,.
\end{equation}

\subsection{(Massive) type IIA $f(R)$ supergravity}

Type IIA supergravity \cite{campwest} is the low-energy effective
theory for the type IIA string.  In addition to the NS-NS fields
that are also present in ${\cal N}=1$, $D=10$ supergravity, the
bosonic sector includes the R-R vector ${\cal A}_\1$ and tensor
$A_\3$ as well. The bosonic Lagrangian is
\begin{equation}
e^{-1} {\cal L}_{10} = R - \ft12 (\partial\phi)^2 - \ft12 e^{-\phi}
F_\3^2 - \ft1{48} e^{\fft12\phi} F_\4^2 - \ft14 e^{\fft32\phi} {\cal
F}_\2^2 + e^{-1} {\cal L}_{\rm FFA}\,.\label{type2aboslag}
\end{equation}
where
\begin{equation}
{\cal F}_\2=d{\cal A}_\1\,,\qquad F_\3 = dA_\2\,,\qquad F_\4=dA_\3 +
A_\1\wedge {\cal F}_\2\,.
\end{equation}
Choosing the $\kappa=+1$ conformal scaling, we have
\begin{equation}
e^{-1} {\cal L}_{10} = \varphi (R -\ft1{48} F_\4^2) - \ft14
\varphi^3 F_\2^2 - \ft1{12} \varphi^{-1}
F_\3^2\,.\label{type2alagscale}
\end{equation}
Thus we see that the $f(R)$ frame is the same as that of the
D2-branes. In order to find $f(R)$, we can first find $F(R)$, which
satisfies the following polynomial
\begin{equation}
36 F_\2^2 F^4 + (F_\4^2 - 48 R) F^2- 4 F_\3^2=0\,,\label{type2acons}
\end{equation}
The $f(R)$ is thus given by
\begin{eqnarray}
f&=&\ft1{432} (2F_\4^2 + X- 96 R) \sqrt{\fft{X - F_\4^2 + 48
R}{2F_\2^2}}\,,\cr 
X^2 &=& 576 F_\2^2 F_\3^2 + (F_\4^2 - 48 R)^2\,.\label{type2af}
\end{eqnarray}
In general, it is smooth to take various field strength to zero.
This can be seen from (\ref{type2acons}).  A special case arises
when we consider $F_\3=0={\cal F}_\2$, leaving only the $F_\4$
non-vanishing.  In this case, the (\ref{type2alagscale}) implies
that
\begin{equation}
e^{-1} {\cal L}_{10} = \varphi (R - \ft1{48}
F_\4^2)\,,\label{4formlag}
\end{equation}
which corresponding to the singular case discussed in section 2.
From the point of view of (\ref{type2af}), it is a singular limit by
letting $F_\3$ and ${\cal F}_\2$ vanish simultaneously, and the
regularity requires us to introduce a scalar proportional to $F$ in
order to keep the theory regular, leading to (\ref{4formlag}).

     The harmonic electric D2-brane can also be easily constructed.
We shall present only the isotropic solution
\begin{eqnarray}
ds_{10}^2 &=& H^{-\fft23} dx^\mu dx_\mu + H^{\fft13} (dr^2 +
r^2d\Omega_6^2)\,,\cr 
\varphi &=& H^{\fft16}\,,\qquad F_\4=d^3x\wedge dH^{-1}\,,\qquad H=1
+ \fft{q}{r^5}\,.
\end{eqnarray}
It is clear that the solution suffers from a power-law curvature
singularity at $r=0$, as in the case in the original Einstein frame.

      There exists an alternative non-harmonic D2-brane
\begin{eqnarray}
ds_{10}^2 &=& H^{-\fft23} dx^\mu dx_\mu + H^{\fft13} (dr^2 +
r^2d\Omega_6^2)\,,\cr 
\varphi &=& (r^6+q)^{\fft16}\,|\cos\theta|\,,\qquad F_\4=d^3x\wedge
dH^{-1}\,,\qquad H=1 + \fft{q}{r^6}\,,\cr 
d\Omega_6^2 &=& d\theta^2 + \sin^2\theta d\Omega_5^2
\end{eqnarray}
The electric charge is given by
\begin{equation}
Q_2=\int_{r\rightarrow \infty} \varphi F_\4 = \fft{6\omega_7}{\pi}
q\,.
\end{equation}
The discussion of the properties of this solution is analogous to
that of the string solution in the previous subsection, and hence we
shall not elaborate further.

The Lagrangian admits a magnetic D4-brane solution, given by
\begin{eqnarray}
ds_{10}^2 &=& H^{-\fft13} dx^\mu dx_\mu + H^{\fft23} (dr^2 +
r^2d\Omega_4^2)\,,\cr 
\varphi &=& H^{-\fft16}\,,\qquad F_\4=3q\Omega_\4\,,\qquad H=1 +
\fft{q}{r^3}\,.
\end{eqnarray}
where $\Omega_\4$ is the volume form for the unit $S^4$.  Thus the
solution interpolates between AdS$_6\times S^4$ at the horizon $r=0$
to the asymptotic flat spacetime at $r=\infty$. The quantity
$F_\4^2$ is clearly finite in this region.  Furthermore, the scalar
$\varphi$ is also finite, running from zero 0 to 1.  From the point
of view of $f(R)$ theory, the $\varphi$ is simply $F$ and hence the
whole solution should be viewed as regular.  This interpretation is
very different from the D4-brane in the usual type IIA supergravity.
Note that here $H$ is the harmonic function in the transverse space,
and hence solution can be generalized to describe multi-center
D4-branes. The AdS$_6\times S^4$ solution can be obtained by
dropping the ``1'' in the function $H$, namely
\begin{eqnarray}
ds_{10}^2 &=& \ell^2 \Big(\fft{dr^2}{r^2} + \ell^{-3} r dx^\mu
dx_\mu + d\Omega_\4^2\Big)\,,\cr 
F_\4 &=& 3\ell^3\Omega_\4\,,\qquad F=\sqrt{\fft{r}{\ell}}
\end{eqnarray}
Note that the proper coordinate for the AdS space is $r=\ell
\rho^2$.  It follows that $r$ will not go negative.  In terms of the
coordinate $\rho$, the D4-brane metric can be viewed as a symmetric
worm-3brane separated by a bulk horizon.  The positiveness of
$\varphi$ requires that $\varphi\equiv F=|\rho|$ and hence the
solution requires a delta function source.  This is analogous to the
type IIB instanton solution \cite{ggpinstanton}, which is a wormhole
in the string frame. The Einstein equations of motion however
involves $\Box H$ where $H$ is the harmonic function on
ten-dimensional Euclidean space.  Thus a matter source is also
needed.

     Lifting the D4-brane to $D=11$ gives rise to the standard
isotropic M5-brane, in which case the solution is totally regular
and there is no need for any source.  This is because only
$r=\rho^2$ appears in the metric and hence the sign choice of $\rho$
is irrelevant.  For this reason one can identify the inside with the
outside of the M5-brane \cite{ghtres}.  The different interpretation
suggests that the $f(R)$ type IIA supergravity contains states that
are outside the physical spectrum in M-theory.

The NS-NS string and 5-brane were already presented in the previous
subsectoin. D0-brane and D6-brane solutions are given by
\begin{eqnarray}
D0:&& ds^2 = -H^{-1} dt^2 + dr^2 + r^2 d\Omega_8^2\,,\cr 
&& {\cal F}_\2 = dt\wedge dH^{-1}\,,\qquad H=1 + \fft{q}{r^7}\,,\cr
D6:&& ds^2 = dx^\mu dx_\mu + H (dr^2 + r^2 d\Omega_\2^2)\,,\cr 
&& {\cal F}_\2 = q\Omega_\2\,,\qquad H=1 + \fft{q}{r}\,.
\end{eqnarray}
Both solutions are singular at $r=0$.

Massive type IIA supergravity in ten dimensions were constructed in
\cite{massive2a}.  After the conformal scaling
(\ref{conformscaling}), the bosonic Lagrangian is given by
\begin{equation}
e^{-1} {\cal L}_{10} = \varphi (R -\ft1{48} F_\4^2) -\ft1{12}
\varphi^{-1} F_\4^2 - \ft14 \varphi^3 {\cal F}_\2^2 -\ft12 m^2
\varphi^5 + e^{-1} {\cal L}_{FFA}\,,
\end{equation}
where
\begin{eqnarray}
{\cal F}_\2 &=& dA_\1 + m A_\2\,,\qquad F_\3=d A_\2\,,\cr 
F_\4 &=& dA_\3 + {\cal A}_\1\wedge dA_\2 + \ft12 A_\2\wedge
A_\2\,,\cr 
{\cal L}_{\rm FFA} &=& \Sigma_{\sst{(10)}}\,,\qquad
\hbox{with}\qquad d\Sigma_{\sst{(10)}}=-\ft12 F_\4\wedge F_\4\wedge
F_\3\,.
\end{eqnarray}
The essence of this theory is that the NS-NS 2-form eats the R-R
vector fields and becomes massive. The $\varphi\equiv F(R)$ function
satisfies the following polynomial equation
\begin{equation}
\ft52m^2 F^6 + \ft34 {\cal F}_\2^2 F^4 + (\ft1{48} F_\4^2 - R) F^2 +
\ft1{12} F_\3^2 =0\,.
\end{equation}
Whilst the analytic form for $F$ exists in this case, it is not
instructive to give explicitly here.  One extra ingredient in
massive type IIA theory is the exponential scalar potential
associated with the D8-brane, where $m$ is the 8-brane change.  The
corresponding $f(R)$ description is
\begin{equation}
f(R)=\fft45 \Big(\fft{2}{5m^2}\Big)^{\fft14}\, R^{\fft54}\,,\qquad
F(R)=\Big(\fft{2}{5m^2}\Big)^{\fft14} R^{\fft14}\,.
\end{equation}
Note that we have also presented the $F(R)$ here.  Thus for this
$f(R)$ theory, any metric with vanishing Ricci scalar $R$ is a
solution, since $f(0)=0=F(0)$.  Since the equations of motion is
reduced to a scalar-like equation $R=0$, there is no propagating
spin-2 degrees of freedom in this background. (See {\it
e.g.}~\cite{llwfr}.) This solution does not exist in the original
massive type IIA supergravity and it cannot be lifted to $D=11$
either.

\subsection{Type IIB $f(R)$ supergravity}

Type IIB supergravity was constructed in \cite{schwarz} at the level
of equations of motion since there can be no Lagrangian formalism
for the self-dual 5-form field strength.  It is nevertheless
possible to write a Lagrangian with a non-self-dual 5-form and then
impose the self-duality by hand after deriving the equations of
motion from the Lagrangian \cite{bho}. After the conformal scaling
(\ref{conformscaling}), the bosonic Lagrangian of the type IIB
supergravity becomes
\begin{eqnarray}
e^{-1} {\cal L} &=& \varphi R - \ft1{12} \varphi^{-1} (F^{\rm
NS}_\3)^2 -\ft12 \varphi^4 (\partial\chi)^2 - \ft1{12} \varphi^2
(F_\3^{\rm RR})^2 - \ft{1}{240} F_\5^2 + e^{-1}{\cal L}_{\rm FFA}\,,
\end{eqnarray}
where
\begin{eqnarray}
&& F_\3^{\rm NS} = dA_\2^{\rm NS}\,,\qquad F_\3^{\rm RR} =
dA_\2^{\rm RR} - \chi dA_\2^{\rm NS}\,,\cr 
&& dF_\5=F_\3^{\rm NS}\wedge F_\3^{\rm RR}\,,\qquad {\cal L}_{FFA}=
\ft12 A_\4\wedge dA_\2^{\rm NS} \wedge dA_\2^{\rm RR}\,.
\end{eqnarray}
Here we adopt the notation of \cite{lpt}.  The self-duality
condition for the 5-form should be imposed at the level of equations
of motion \cite{bho}. Unfortunately, there can be no analytical
$f(R)$ description of type IIB supergravity since the function $F$
satisfies an equation of the quintic-order polynomials. The $f(R)$
theory for the NS-NS 3-form was given earlier.  The one for each
individual R-R fields will be given in the next subsection. Here we
shall draw attention to the fact that the $f(R)$ theory for the R-R
3-form is identical to that in (\ref{kappaminusfr}), which has an
origin from $D=11$ as the $\kappa=-1$ reduction.

\subsection{Further discussions on the R-R sector}

The D$(p-1)$-branes in supergravities are brane solutions supported
by the R-R $p$-form field strength.  In general, the relevant
bosonic Lagrangian, under the conformal scaling
(\ref{conformscaling}), is
\begin{equation}
e^{-1} {\cal L} = \varphi R - \fft1{2\,p!} \varphi^{5-p} F_p^2\,.
\end{equation}
The associated $f(R)$ is
\begin{equation}
f(R) = \Big(\fft{2\, p!}{5-p}\Big)^{\fft{1}{4-p}}\, \fft{p-4}{p-5}
\,R\, \Big(\fft{R}{F_p^2}\Big)^{\fft{1}{4-p}}\,,\qquad
\hbox{with}\qquad f=\fft{p-4}{p-5}\, R\, F\,.
\end{equation}
Thus the $f(R)$ description appears to break down for $p=4$ and $5$,
corresponding to D4, D6 and D5 branes.  However, this only implies
that when we take a limit to $p=4$ and $5$ cases from the full
general $f(R)$ theory, a proper care should be taken.  Let us
consider the D4-D8 system as an example, which involves both the
$F_\4$ and the scalar potential in the massive type IIA
supergravity. The relevant Lagrangian is
\begin{equation}
e^{-1} {\cal L} = \varphi (R - \ft1{48} F_\4^2) - \ft12 m^2
\varphi^5\,.\label{d4d8lag}
\end{equation}
The $f(R)$ theory is given by
\begin{equation}
f(R)=\ft45 F(R)\, (R - \ft1{48} F_\4^2)\,,\qquad F=\Big(\ft{2}{5m^2}
(R - \ft1{48} F_\4^2)\Big)^{\fft14}\,.
\end{equation}
The second equation above implies that there is a smooth limit with
$m\rightarrow 0$ that recovers (\ref{d4d8lag}) modulo an overall
factor $\ft45$.

It was shown that \cite{boads6,clpd6gauge} AdS$_6$ arises in the
localized D4/D8-brane system \cite{youm}, and the gauged AdS$_6$
supergravity \cite{romans} can be obtained from spherical reduction
from massive type IIA supergravity \cite{clpd6gauge}.  The relevant
$f(R)$ theory for the D4/D8 system is given by above.  The AdS$_6$
embedding embedding \cite{boads6,clpd6gauge} in the $f(R)$ theory
becomes
\begin{eqnarray}
ds^2_{10} &=& (\cos\theta)^{\fft29} ( ds_{AdS_6}^2 + 2 d\theta^2 + 2
\sin^2\theta d\Omega_3^2)\,,\cr 
F_\4 &=& \ft{5\sqrt2}6 (\cos\theta)^{\fft13} \sin^3\theta\,
d\theta\wedge \Omega_\3\,,\qquad F=
(\cos\theta)^{-\fft59}\,.\label{d4d8}
\end{eqnarray}
Thus the solution has a power-law curvature singularity at the
equator $\theta=\pi/2$ of the $S^4$.  The solution becomes regular
in the D4-brane frame \cite{clprotstr}.  In the D4-brane, the
Lagrangian for massive type IIA supergravity has the form
\begin{equation}
e^{-1}{\cal L} = \tilde \varphi (R -2(\partial\log\tilde\varphi)^2 -
\ft{1}{1440} F_\6^2) + \hbox{more}\,.
\end{equation}
The metric of the D4/D8 solution is then simply the AdS$_6\times
S^4$ without the pre-factor in (\ref{d4d8}) and furthermore,
$\tilde\varphi=(\sin\theta)^{1/3}$.  Thus it is necessary that
$\theta$ runs from 0 to $\pi$ with the $(\cos\theta)^{1/3}$ factor
in both $F_\4$ and $\tilde \varphi$ added an absolute-value sign,
namely $|(\cos\theta)^{1/3}|$. This requires a delta-function source
on the equator.  This is very analogous to the new solutions of
ten-dimensional $f(R)$ supergravities we have obtained.

     Before we end this section we would like to remark that in
all supergravities in ten dimensions, the algebraic equation
associated with the dilaton $\varphi$ in the $f(R)$ frame are
polynomials of integer power.  This property is not universal.  In
section 8, we present an example of ${\cal N}=1$, $D=7$ gauged
supergravity, and the resulting polynomial is of irrational powers.
This suggests that ten-dimensional supergravities are special from
the point of view of the $f(R)$ formalism.  Indeed ten-dimensional
supergravities play two roles in string and M-theory. One is that
they are the low-energy effective theories of string; the other is
that they are related to M-theory through the dimensional
compactification.

\section{$f(R)$ supergravities: the fermionic sector}
\label{fermionsection}

In the previous section, we demonstrate that the bosonic sectors of
supergravities in ten dimensions can be converted to $f(R)$ theories
coupled to the form fields.  In this section, we shall consider the
fermion fields and show that such a conversion extended to the
fermionic sector, at least up to the quadratic order in fermions.
The key point is that the dilaton in the $f(R)$ frame remains
auxiliary even when the fermionic sector is included.

\subsection{A general discussion}

All the supergravity theories considered in the previous section
contain a dilaton that is part of the supergravity multiplet. The
truncated Lagrangian involving the metric, dilaton, gravitino and
dilatino takes the universal form
\begin{equation}
e^{-1} {\cal L}_D = R - \ft12(\partial\phi)^2 +\ft12 \bar\psi_\mu
\Gamma^{\mu\nu\rho} D_\nu \psi_\rho + \ft12 \bar \lambda \Gamma^\mu
D_\mu \lambda + \ft{\im}{2\sqrt{2\beta}} \bar\lambda\Gamma^\mu
\Gamma^\nu \psi_\mu \partial_\nu \phi\,,\label{gensusylag0}
\end{equation}
where $\beta=\pm1$ depending on the fermion convention presented in
Table 1 below. (This $\beta$ has nothing to do with the constant
defined in (\ref{alphabeta}).) The relevant parts of the
supersymmetric transformation rules are given by
\begin{eqnarray}
&&\delta\psi_\mu = D_\mu \epsilon\,,\qquad \delta\lambda = \ft{\im
\sqrt{\beta}}{2\sqrt2}\,\Gamma^\mu \partial_\mu \phi \epsilon\,, \cr
&& \delta e^{a}_\mu=\ft14 \bar \psi_\mu \Gamma^a \epsilon\,,\qquad
\delta \phi = \ft{\rm i}{2\sqrt{2\beta}} \bar \epsilon\lambda\,.
\end{eqnarray}
It should be pointed out that the Lagrangian (\ref{gensusylag0}) is
already invariant under the supersymmetric transformation rules up
to the quadratic order in fermions, and hence the theory can be
viewed as pseudo-supergravity discussed in
\cite{lpwpseudo,llwpseudo}.

      We first make the following conformal transformation and field
redefinition,
\begin{eqnarray}
e^a_{\mu} &=& \varphi^{\fft{1}{D-2}}\tilde e^a_\mu\,,\qquad
\psi_\mu=\varphi^{\fft{1}{2(D-2)}} \Big(\tilde \psi_\mu -
\ft{\im}{\sqrt{(D-1)(D-2)\beta}} \Gamma_\mu \tilde\lambda\Big)\,,\cr
\lambda &=&\varphi^{-\fft{1}{2(D-2)}}\tilde\lambda\,,\qquad
\epsilon=\varphi^{\fft{1}{2(D-2)}}\tilde\epsilon\,,\label{fermiontrans}
\end{eqnarray}
where $\varphi$ is related to $\phi$ by (\ref{conformscaling}) and
(\ref{alphabeta}) with $\kappa=+1$.  Substituting these into
(\ref{gensusylag0}) and then dropping the tilde, we find that the
Lagrangian becomes
\begin{eqnarray}
e^{-1} {\cal L}_D &=& \varphi \widehat R\,,\qquad  \widehat R=R +
K\,,\cr 
K &=& \ft12 \bar \psi_\mu \Gamma^{\mu\nu\rho} D_\nu \psi_\rho - \im
\sqrt{\ft{(D-2)\beta}{D-1}}\, \bar \lambda \Gamma^{\mu\nu}
D_\mu\psi_\nu + \ft12 D^\nu(\bar\psi_\mu \Gamma^\mu \psi_\nu) \,.
\end{eqnarray}
The supersymmetric transformation rules become
\begin{eqnarray}
&&\delta\psi_\mu = D_\mu \epsilon\,,\qquad \delta\lambda = \ft{\im
\sqrt{(D-1)\beta}}{2\sqrt{D-2}}\,\varphi^{-1}\Gamma^\mu \partial_\mu
\varphi \epsilon\,, \cr 
&& \delta e^{a}_\mu=\ft14 \bar \psi_\mu \Gamma^a \epsilon\,,\qquad
\delta \varphi = -\ft{\rm i\sqrt{(D-2)\beta}}{4\sqrt{D-1}}\, \bar
\lambda\epsilon\,.
\end{eqnarray}
Note that in converting the theory to the new frame, $\delta
e^a_\mu$ will generate a local Lorentz transformation which we have
dropped.   Thus we see that in the new $f(R)$ frame where the
dilaton loses its kinetic term, the contribution of the fermions is
simply to replace Ricci scalar with $\widehat R$, with $\varphi$
remaining auxiliary. Thus we have the generic result that for this
contribution: we can simply replace the previous $f(R)$ with
$f(\widehat R)$.

\subsection{${\cal N}=1$, $D=10$ $f(R)$ supergravity}

The full ${\cal N}=1$, $D=10$ supergravity theory was constructed in
\cite{bvn}.  In this paper, we shall consider the fermionic sector
only up to the quadratic order in fermion fields. The supersymmetric
partners are gravitino $\psi_\mu$ and dilitino $\lambda$, which are
both Majorana and Weyl.  Making the field redefinition
(\ref{fermiontrans}) and then drop the tilde, we find that that the
Lagrangian in the $f(R)$ frame is given by
\begin{equation}
e^{-1} {\cal L} = \varphi (R+K)  - \ft1{12} \varphi^{-1} F_\3^2 +
X_3,
\end{equation}
where $K$ and the Yukawa term $X_3$ associated with $F_\3$ are given
by
\begin{eqnarray}
K &=& \ft12 \bar\psi_\mu \Gamma^{\mu\nu\rho} D_\nu \psi_\rho -
\ft{2\sqrt2\,\im}3 \bar \lambda \Gamma^{\mu\nu} D_\mu \psi_\nu +
\ft12 D^\mu (\psi_\nu \Gamma^\nu \psi_\mu) \,,\cr 
X_3 &=& \Big(-\ft1{48} \bar \psi_\mu
\Gamma^{\mu\nu\rho\sigma\lambda} \psi_{\lambda} - \ft18
\psi^\nu\Gamma^\rho \psi^\sigma + \ft{\sqrt2\,\im}{12} \bar \lambda
\Gamma^{\nu\rho} \psi^\sigma\Big) F_{\nu\rho\sigma}\,.
\end{eqnarray}
The supersymmetric transformation rules in the $f(R)$ frame are
given by
\begin{eqnarray}
\delta\psi_\mu &=& D_\mu \epsilon + \ft{1}{54} F_{\nu\rho\sigma}
\Gamma_\mu{}^{\nu\rho\sigma} \epsilon- \ft1{12} F_{\mu\nu\rho}
\Gamma^{\nu\rho} \epsilon\,,\cr 
\delta \lambda &=& \ft{3\im}{4\sqrt2} \varphi^{-1} (\Gamma^\mu
\partial_\mu \varphi- \ft1{18} F_{\mu\nu\rho}
\Gamma^{\mu\nu\rho})\epsilon\,,\cr 
\delta e^{a}_\mu &=& \ft14\bar\psi_\mu\Gamma^a \epsilon\,,\qquad
\delta g_{\mu\nu} = \ft12 \bar \psi_{(\mu} \gamma_{\nu)}
\epsilon\,,\qquad \delta \varphi = -\ft{\im}{3\sqrt2} \varphi
\bar\lambda \epsilon\,,\cr 
\delta A_{\mu\nu} &=& \varphi (-\ft12 \bar \epsilon \Gamma_{[\mu}
\psi_{\nu]} + \ft{\im}{3\sqrt2} \bar \epsilon \Gamma_{\mu\nu}
\lambda)
\end{eqnarray}
Integrating out $\varphi$, we find that the $f(R)$ theory of the
${\cal N}=1$, $D=10$ $f(R)$ supergravity is given by
\begin{eqnarray}
e^{-1} {\cal L} &=& \sqrt{-\ft13 (R+K) F_\3^2} + X_3\,,\cr 
\delta\psi_\mu &=& D_\mu \epsilon + \ft{1}{54} F_{\nu\rho\sigma}
\Gamma_\mu{}^{\nu\rho\sigma} \epsilon- \ft1{12} F_{\mu\nu\rho}
\Gamma^{\nu\rho} \epsilon\,,\cr 
\delta \lambda &=& \ft{3\im}{4\sqrt2} F^{-1} (\Gamma^\mu
\partial_\mu F- \ft1{18} F_{\mu\nu\rho}
\Gamma^{\mu\nu\rho})\epsilon\,,\cr 
\delta e^{a}_\mu &=& \ft14\psi_\mu\Gamma^a \epsilon\,,\qquad \delta
g_{\mu\nu} = \ft12 \bar \psi_{(\mu} \gamma_{\nu)} \epsilon_0\,,\cr
\delta A_{\mu\nu} &=& F\, (-\ft12 \bar \epsilon \Gamma_{[\mu}
\psi_{\nu]} + \ft{\im}{3\sqrt2} \bar \epsilon \Gamma_{\mu\nu}
\lambda)
\end{eqnarray}

One can add further the matter Yang-Mills multiplet $(A_\mu, \chi)$.
In the $f(R)$ frame, the extra parts of the Lagrangian and
supersymmetric transformation rules are given by
\begin{eqnarray}
e^{-1}{\cal L}_{\rm YM} &=& -\ft14 F_\2^2 +\ft12 \varphi\bar\chi
\Gamma^\mu D_\mu \chi -\ft1{48} F_{\mu\nu\rho} \bar\chi
\Gamma^{\mu\nu\rho} \chi -\ft{1}{4\sqrt2} \varphi^{\fft12}
F_{\nu\rho} \bar\chi \Gamma^\mu \Gamma^{\nu\rho} \psi_\mu\,,\cr 
\delta \chi &=& \ft1{4\sqrt2} \Gamma^{\mu\nu} F_{\mu\nu}
\epsilon\,,\qquad \delta A_\mu = -
\ft1{2\sqrt2}\varphi^{\fft12}\bar\chi \Gamma_\mu \epsilon\,,\cr 
\delta_{\rm extra} A_{\mu\nu} &=& \ft1{2\sqrt2} \varphi^{\fft12}
\bar\chi A_{[\mu} \Gamma_{\nu]} \epsilon\,.
\end{eqnarray}
It is again straightforward to integrate out the auxiliary $\varphi$
and obtain the $f(R)$ theory of heterotic supergravity.

\subsection{Type IIA $f(R)$ supergravity}

     Type IIA supergravity was constructed in \cite{campwest}. The
superpartners include the Majorana gravitino and dilatino. We first
convert the the mostly minus convention of the spacetime signature
of \cite{campwest} to the mostly plus one. Making an analogous field
reduction (\ref{fermiontrans}) with appropriate $\Gamma_{11}$
inserted, we find that the Lagrangian of type IIA supergravity in
the $f(R)$ frame up to the quadratic fermionic order is given by
\begin{eqnarray}
e^{-1}{\cal L}_{10}\!\!\!&=&\!\!\!e^{-1} {\cal L}_{\rm B} + \varphi
\Big(\ft12 \bar\psi_\mu \Gamma^{\mu\nu\rho} D_\nu \psi_\rho -
\ft{2\sqrt2\,\im}3 \bar \lambda \Gamma^{\mu\nu}\Gamma_{11} D_\mu
\psi_\nu + \ft12 D^\mu (\psi_\nu \Gamma^\nu \psi_\mu)\Big)\cr 
\!\!\!&&\!\!\!+\Big(-\ft1{48} \bar \psi_\mu\Gamma_{11}
\Gamma^{\mu\nu\rho\sigma\lambda} \psi_{\lambda} - \ft18
\psi^\nu\Gamma_{11}\Gamma^\rho \psi^\sigma + \ft{\sqrt2\,\im}{12}
\bar \lambda\Gamma_{11}\Gamma^{\nu\rho} \psi^\sigma\Big)
F_{\nu\rho\sigma}\cr 
\!\!\!&&\!\!\!+\varphi^2\Big(
\ft{1}{16}\bar\psi_{\alpha}\Gamma^{\alpha\beta\mu\nu} \psi_{\beta}
+\ft{1}{8}\bar\psi^{\mu}\psi^{\nu}+ \ft{\sqrt{2}}{24} \bar \lambda
\Gamma^{\alpha} \Gamma^{\mu\nu}\psi_{\alpha}
+\ft{\sqrt{2}}{6}\bar\lambda \Gamma^{\mu} \psi^{\nu}
-\ft{1}{144}\bar \lambda \Gamma_{11} \Gamma^{\mu\nu} \lambda
\Big)F_{\mu\nu}\cr 
\!\!\!&&\!\!\!+\varphi \Big(-\ft{1}{192}\bar
\psi_{\alpha}\Gamma^{\alpha\mu\nu\rho\sigma\beta}
\psi_{\beta}-\ft{1}{16}\bar\psi^{\mu} \Gamma^{\nu\rho}
\psi^{\sigma}+\ft{1}{72\sqrt{2}}\bar \lambda
\Gamma_{11}\Gamma^{\mu\nu\rho\sigma\alpha} \psi_{\alpha}\Big)
F_{\mu\nu\rho\sigma}\,.
\end{eqnarray}
where ${\cal L}_{\rm B}$ is given in (\ref{type2alagscale}). We see
that at least up to the quadratic fermion order, the scalar
$\varphi$ is auxiliary and can be integrated out, yielding an $f(R)$
theories of type IIA supergravity. Furthermore, there is no
derivative on the dilatino $\lambda$ in the Lagrangian and hence it
is also an auxiliary field and can be integrated out.

    The supersymmetric transformation rules in the $f(R)$ frame
up to quadratic order in fermions are given by
\begin{eqnarray}
&&\delta\psi_\mu = D_\mu \epsilon - \Big(\ft{1}{288}
\Gamma_\mu{}^{\nu\rho\sigma\lambda} - \ft1{36}\delta_{\mu}^\nu
\Gamma^{\rho\sigma\lambda}\Big) F_{\nu\rho\sigma\lambda} \epsilon -
\varphi \Big(\ft1{32} \Gamma_\mu{}^{\nu\rho} - \ft3{10}
\delta_\mu^\nu \Gamma^\rho\Big)\Gamma_{11} F_{\nu\rho} \epsilon\cr
&&\qquad\qquad-\varphi^{-1}\Big(\ft1{72}
\Gamma_{\mu}{}^{\nu\rho\sigma}-\ft1{12}
\delta_{\mu}^{\nu}\Gamma^{\rho\sigma}\Big)\Gamma_{11}
F_{\nu\rho\sigma} \epsilon\,,\cr 
&&\delta\lambda = - \ft{3}{4\sqrt2} \Big( \Gamma^\mu\Gamma_{11}
\varphi^{-1} \partial_\mu \varphi - \ft{1}{144}
\Gamma^{\mu\nu\rho\sigma} \Gamma_{11} F_{\mu\nu\rho\sigma} -
\ft{1}{18} \varphi^{-1} \Gamma^{\mu\nu\rho} F_{\mu\nu\rho} +
\ft14\varphi \Gamma^{\mu\nu} F_{\mu\nu}\Big)\epsilon\,,\cr 
&&\delta e^{a}_\mu = \ft14\bar\psi_\mu\Gamma^a \epsilon\,,\qquad
\delta g_{\mu\nu} = \ft12 \bar \psi_{(\mu} \gamma_{\nu)}
\epsilon\,,\qquad \delta \varphi = -\ft{1}{3\sqrt2} \varphi
\bar\lambda \Gamma_{11}\epsilon\,,\cr 
&&\delta A_\mu = -\ft14 \varphi^{-1} \Big(\bar \psi_\mu \Gamma_{11}
- \ft{2\sqrt2}{3} \bar \lambda \Gamma_\mu\Big)\epsilon \,,\cr 
&& \delta A_{\mu\nu}=-\ft12\varphi\Big(\bar\psi_{[\mu}\Gamma_{\nu]}
- \ft{\sqrt2}{3} \bar\lambda \Gamma_{\mu\nu}\Big)\epsilon\,,\cr 
&&\delta A_{\mu\nu\rho}=-\ft34 \bar\psi_{[\mu} \Gamma_{\nu\rho]}
\epsilon + 3A_{[\mu}\delta A_{\nu\rho]}\,.
\end{eqnarray}

\section{$f(R)$ gauged (pseudo) supergravities}

In this section, we demonstrate that some gauged supergravities can
be converted into $f(R)$ supergravities.  We then demonstrate that
the $f(R)$ theories admit new solutions with global properties that
do not exist in the corresponding solutions of the original
theories.

\subsection{${\cal N}=2$, $D=5$ gauged $f(R)$ supergravity}

The minimum supergravity in $D=5$ has ${\cal N}=2$ supersymmetry
from the counting scheme in $D=4$. It contains the metric, a Maxwell
vector and a gravitino that is a sympletic majorana.  The theory can
be gauged with the sympletic structure broken down to the $U(1)$
symmetry. (See {\it e.g.} \cite{chni}.) The theory with an arbitrary
number of vector multiplet was constructed in \cite{gst}. Here, we
shall consider only one vector multiplet which consists of a scalar,
a vector and a dilatino.  The Lagrangian for the bosonic sector is
given by
\begin{eqnarray}
e^{-1}{\cal L}_5 &=& R - \ft12(\partial \phi)^2 +4g^2 \Big(2
e^{-\fft{1}{\sqrt6} \phi} + e^{\fft2{\sqrt6} \phi}\Big) -
\ft14e^{-\fft{2}{\sqrt6}\phi} F_\2^2 - \ft14 e^{\fft4{\sqrt6}\phi}
{\cal F}_\2^2\cr 
&& +\ft18 e^{-1}\epsilon^{\mu\nu\rho\sigma\lambda} F_{\mu\nu}
F_{\rho\sigma} {\cal A}_\lambda\,,\label{d5suplag}
\end{eqnarray}
where $F_\2=dA_\1$ and ${\cal F}_\2=d{\cal A}_\1$. The
supersymmetric transformation rules are given by
\begin{eqnarray}
\delta\psi_\mu &=& [D_\mu - \ft{\rm i}2 g (\sqrt2 A_\mu + {\cal
A}_\mu)]\epsilon + \ft16 g (2e^{\fft1{\sqrt6}\phi} +
e^{-\fft2{\sqrt6}\phi}) \epsilon\cr 
&& + \ft{\rm i}{8} (\Gamma_{\mu}{}^{\nu\rho} - 4 \delta_{\mu}^\nu
\Gamma^\rho) ( \sqrt2 e^{-\fft1{\sqrt6}\phi} F_{\nu\rho} +
e^{\fft2{\sqrt6}\phi}{\cal F}_{\nu\rho})\Gamma_\mu\epsilon\,,\cr 
\delta\lambda &=& -\ft{\rm i}4 \Gamma^\mu\partial_\mu \phi \epsilon
+ \ft{\rm i}{\sqrt6} g (e^{\fft{1}{\sqrt6}\phi} -
e^{-\fft2{\sqrt6}\phi} )\epsilon\cr 
&&- \ft3{8\sqrt6} (\sqrt2 e^{-\fft1{\sqrt6}\phi} F_{\mu\nu} -
2e^{\fft2{\sqrt6}\phi} {\cal F}_{\mu\nu})\Gamma^{\mu\nu}\epsilon\,,
\end{eqnarray}
We have also consulted the papers \cite{bcs1,bcs2,tenauthor} in
deriving the above results. If we set $g=0$, the theory becomes the
ungauged ${\cal N}=2$ supergravity in $D=5$ with a vector multiplet.
The ungauged theory can be obtained from the $S^1$ reduction
\cite{stainless} of the ${\cal N}=(1,0)$, $D=6$ supergravity.  The
vector ${\cal A}_\1$ is simply the Kaluza-Klein vector, and $A_\1$
has an origin of the self-dual 3-form field strength in $D=6$.

Under the conformal scaling (\ref{conformscaling}), the Lagrangian
(\ref{d5suplag}) becomes
\begin{equation}
e^{-1} {\cal L} = \varphi (R + 8g^2) + \varphi^3 (4g^2 - \ft14 {\cal
F}_\2^2) - \ft14 \varphi^{-1} F_\2^2 - \ft14 e^{-1}
\epsilon^{\mu\nu\rho\sigma\lambda} F_{\mu\nu} F_{\rho\sigma} {\cal
A}_\lambda\,.\label{d5suplag1}
\end{equation}
Note that apart from the $FFA$ and $F_\3$ term, this Lagrangian is
the same as (\ref{gaugegenfr}) in $D=5$. It should be reminded that
it is typical that the Lagrangians of supergravities in the $f(R)$
frame are polynomials of $\varphi$ with irrational powers. Whilst
all ten-dimensional supergravities have integer powers, such
exceptions are rare in lower dimensions.

Performing the following field redefinition for the fermions:
\begin{equation}
\psi_\mu = \varphi^{-\fft16} (\tilde\psi_\mu + \ft{\im}{\sqrt6}
\Gamma_\mu\tilde \lambda)\,,\qquad \lambda = \varphi^{-\fft16}
\tilde\lambda\,,\qquad \epsilon=\varphi^{\fft16}\tilde\epsilon\,,
\end{equation}
we find that the supersymmetric transformation rules, after dropping
the tilde, become
\begin{eqnarray}
\delta\psi_\mu &=& [D_\mu - \ft{\im}{2} g ( \sqrt2 A_\mu + {\cal
A}_\mu)] \epsilon + \ft13 g \varphi \Gamma_\mu \epsilon\cr 
&&+ \ft{3\sqrt2\,\im}{16} \varphi^{-1} F_{\nu\rho} \Gamma_\mu
\Gamma^{\nu\rho}\epsilon -\ft{3\im}{4} (\sqrt2\, \varphi^{-1}
F_{\mu\nu} + \varphi {\cal F}_{\mu\nu}) \Gamma^\nu\epsilon\,,\cr 
\delta\lambda &=& -\fft{\im}{\sqrt6} \Big( \varphi^{-1}
\Gamma^\mu\nabla_\mu \varphi - g (\varphi-\varphi^{-1}) -
\ft{3\im}{8}( \sqrt2\varphi^{-1} F_{\mu\nu} - 2\varphi {\cal
F}_{\mu\nu}) \Gamma^{\mu\nu}\Big)\epsilon\,.
\end{eqnarray}
Thus the scalar $\varphi$ in (\ref{d5suplag1}) is auxiliary, and can
be integrated out straightforwardly, giving to the $f(R)$ theory of
gauged supergravity. The algebraic equation for $\varphi^2$ is a
second-order polynomial and hence can be solved explicitly.

   Since our $f(R)$ gauged supergravity is derived from the
previously-known ${\cal N}=2$, $D=5$ gauged supergravity with a
vector multiplet, the result is interesting only if the $f(R)$
supergravity is not equivalent to the original theory.  This can be
seen by examining the solution space of the two theories.  Note that
in the absence of the Maxwell fields, the $f(R)$ theory is identical
to the second example discussed in section 2 with $D=5$.  It follows
that this theory admits the following AdS worm-brane solution
\begin{equation}
ds_5^2 = dr^2 + \cosh^2(g r) (-dt^2 + dx_1^2 + dx_2^2 +
dx_3^2)\,,\qquad F=\tanh(g|r|)\,.
\end{equation}
The solution is BPS, preserving $\ft12$ of the supersymmetry, with
Killing spinor given by
\begin{equation}
\tilde \epsilon = \cosh^{\fft12}(gr)\,\epsilon_0\,,\qquad
\Gamma_r\epsilon_0=\epsilon_0\,.\label{d5adsworm}
\end{equation}
One should not view the worm-brane solution (\ref{d5adsworm}) simply
as a conformal scaling of some domain walls in the original theory.
This is because the Ricci scalar $R$ runs from $R=-20g^2$ at the two
$r\rightarrow \pm \infty$ asymptotic boundaries to $R=-8g^2$ at
$r=0$, at which point $F=0$. Since $F$ is the conformal scaling
factor between the two theories, the connection between the two
theories breaks down for solutions with vanishing or divergent $F$.
The absolute-value sign appearing in $F$ implies that some
delta-function matter source is needed for sustaining the wormhole.

One may argue that the power-law curvature singularity of this
solution at $r=0$ of the original gauged supergravity theory is
merely an artefact of dimensional reduction since the theory can be
embedded in type IIB supergravity.  Using the reduction anstaz given
in \cite{tenauthor}, we find that the ten-dimensional metric is
\begin{eqnarray}
ds^2_{10}&=&\sqrt{\Delta}\Big(dr^2 + \cosh^2(g r) (-dt^2 + dx_1^2 +
dx_2^2 + dx_3^2) + g^{-2} d\theta^2\Big)\cr 
&&+ \fft{1}{g^2\sqrt{\Delta}} \Big(\sin^2\theta\, d\Omega_3^2 +
\tanh^2(gr)\, \cos^2\theta\, d\phi^2\Big)\,,\cr 
\Delta &=& \cos^2\theta + \tanh^2(gr)\, \sin^2\theta\,.
\end{eqnarray}
In $D=10$, the coordinate $r$ runs from $\infty$ to 0, without
having to go to the negative values of $r$. This is because when
$r\rightarrow 0$,  the metric becomes
\begin{equation}
ds_{10}^2=\cos\theta \Big(dr^2 + r^2 d\phi^2 + dx^\mu dx_\mu +
g^{-2} d\theta^2 + g^{-2} \tan^2\theta\, d\Omega_3^2\Big)\,.
\end{equation}
The solution does have a power-law curvature singularity when $r$
and $\cos\theta$ vanish simultaneously.  Thus we see that the
solution behaves differently in $D=11$ and in $D=5$.  In the $D=5$
$f(R)$ theory, it is natural for the coordinate $r$ to run from
$-\infty$ to $+\infty$, whilst in $D=10$, it runs from 0 to
$\infty$.

      In \cite{bcs2}, non-extremal black hole solutions of the
$U(1)^3$ theory in $D=5$ was constructed.  Setting the two of the
three vectors equal gives rise to the theory we are discussing in
this section.  In the BPS limit, the solutions \cite{bcs1} suffer
from a naked power-law curvature singularity and hence are not black
holes. They are often referred as ``superstars.'' We now consider
these superstars, whose generalization to arbitrary dimensions were
given in section 3.3.  If we turn on only the $F_\2$ charge, the
solution of the $f(R)$ theory is given by
\begin{eqnarray}
ds_5^2 &=& - H^{-1} h dt^2 + H \Big(\fft{dr^2}{h} + r^2
d\Omega_3^2\Big)\,,\qquad F_\2=\sqrt2\, dt\wedge dH^{-1}\,,\cr 
h &=& 1 + g^2 r^2 H^2\,,\qquad F =\fft{|r|}{\sqrt{r^2 +
q^2}}\,,\qquad H=1 + \fft{q}{r^2}\,.
\end{eqnarray}
Since we have
\begin{equation}
H^{-1} h = \fft{r^2 + g^2 (r^2 + q^2)^2}{r^2 + q^2}\,,
\end{equation}
it follows that the solution describes a wormhole with $r$ runs from
$-\infty$ to $+\infty$.  The positivity of $F$ requires that a
delta-function matter source at $r=0$ is needed for supporting this
wormhole.  Compared to the previous domain wall solution, this
charged solution is more accurately called a wormhole since the
$r=\rm constant$ slice is $S^3\times \mathbb{R}^{t}$ rather than the
four-dimensional Minkwoski spacetime. (Note that if we turn on both
$F_\2$ and ${\cal F}_\2$, the metric around $r=0$ becomes $ds^2\sim
r^2 dr^2 + \cdots$, implying that $r^2$ can be negative and hence
the solution has a naked singularity when $H$ vanishes.)  The
corresponding $D=10$ type IIB solution is given by
\begin{eqnarray}
ds^2_{10}&=&\sqrt{\Delta}\Big(-\fft{r^2 + g^2 (r^2 + q^2)^2}{r^2 +
q^2} dt^2 + \fft{(r^2 + q^2)dr^2}{r^2 + g^2 (r^2 + q^2)^2} + (r^2 +
q^2) d\Omega_3^2 + g^{-2} d\theta^2\Big)\cr 
&&+ \fft{1}{g^2\sqrt{\Delta}} \Big(\ft14\sin^2\theta\,(\sigma_1^2 +
\sigma_2^2 + (\sigma_3 + 2 A_\1)^2) + \fft{r^2}{r^2+q^2}
\cos^2\theta\, d\phi^2\Big)\,,\cr 
\Delta &=& \cos^2\theta + \fft{r^2}{r^2+q^2} \sin^2\theta\,.
\end{eqnarray}
It is thus clear that the coordinate $r$ runs from $0$ to $\infty$
in $D=10$.

  There is another solution in the $f(R)$ supergravities that does
not exist in the original gauged supergravity.  That is the solution
with $R=-8g^2$.  Since for this case, both $f$ and $F$ vanish, and
hence the full equations of motion is reduced to simply a
scalar-type equation $R=-8g^2$.

\subsection{$D=4$ gauged supergravity}

Four-dimensional maximum gauged supergravity has an $SO(8)$ local
gauge group.  For the bosonic sector, it is consistent to truncate
to the $U(1)^4$ Cartan subsector. (See, {\it
e.g.}~\cite{duffliu,tenauthor}.) We shall consider the special case
where three of the $U(1)$ vectors are set to equal. Following the
results given in \cite{tenauthor}, we find that the Lagrangian is
given by
\begin{equation}
e^{-1}{\cal L}= R - \ft12 (\partial \phi)^2 - \ft14
e^{-\fft1{\sqrt3}\phi} F_\2^2 - \ft14 e^{\sqrt3\phi} {\cal F}_2^2 +
3g^2 (e^{-\fft1{\sqrt3}\phi} +
e^{\fft1{\sqrt3}\phi})\,.\label{d4gaugelag}
\end{equation}
In the $f(R)$ frame, this theory becomes
\begin{equation}
e^{-1} {\cal L}= \varphi (R + 3 g^2) + \varphi^3 (4g^2 - \ft14 {\cal
F}_\2^2) - \ft14 \varphi^{-1} F_\2^2\,,\label{d4gaugefr}
\end{equation}
Thus we see that aside from the $F_\3$ term, this Lagrangian is the
same as the one discussed in section 3.4.  Thus the $f(R)$ admits
both the worm-brane (\ref{wormbrane}) and charged wormhole
(\ref{chargwormhole}) solutions with $D=4$.  Using the reduction
ansatz given in \cite{tenauthor} we can lift the solutions back to
$D=11$. The metric of the smeared M2-brane associated with the
worm-brane is given by
\begin{eqnarray}
ds^2_{11} &=& \Delta^{2/3} \Big(dr^2 + \cosh^4(\ft12g\,r) dx^\mu
dx_\mu + \fft{4d\theta^2}{g^2}\Big)\cr 
&& + \fft{4}{g^2\Delta^{1/3}} \Big(\tanh^2(\ft12 g r) \cos^2\theta
d\phi^2 + \sin^2\theta d\Omega_5^2\Big)\,,\cr 
\Delta&=& \cos^2\theta + \tanh^2(\ft12 gr)\,\sin^2\theta\,.
\end{eqnarray}
The lifting of the charged wormhole solution to $D=11$ gives rise to
\begin{eqnarray}
ds^2_{11} &=& \Delta^{2/3} \Big [-\fft{\rho^2 + g^2(\rho^2 +
q)^3}{\rho^2 + q^2} dt^2 + (\rho^2 + q)^2 \Big(
\fft{4d\rho^2}{\rho^2 + g^2 (\rho^2 + q)^3} +
d\Omega_2^2\Big)\Big]\cr 
&&+\fft{4}{g^2 \Delta^{1/3}} \Big[\fft{\rho^2}{\rho^2 + q}
\cos^2\theta d\phi^2 + \sin^2\theta \Big( (d\psi + \fft{\sqrt3\,
q}{\rho^2 + q^2}dt + 2B_\1)^2 + d\Sigma_{\CP^2}^2\Big)\Big]\,,\cr 
\Delta &=& \cos^2\theta + \fft{\rho^2}{\rho^2 + q} \sin^2\theta\,.
\end{eqnarray}
where $J_\2=dB_\1$ is the K\"ahler 2-form of the $\CP^2$ metric
$d\Sigma^2_{\CP^2}$.  Thus we see that in $D=11$, both coordinates
$r$ and $\rho$ run from 0 to $\infty$, whilst in the $D=4$ $f(R)$
theory, they run from $-\infty$ to $+\infty$.

\subsection{Gauged $f(R)$ Kaluza-Klein pseudo-supergravities}

   It was shown in \cite{llw} that the Lagrangian (\ref{gaugedKKlag})
admits Killing spinor equations in general dimensions. Consequently
the pseudo-supersymmetric extension of the system was constructed in
\cite{llwpseudo}.  In the $f(R)$ frame, we find that the full
Lagrangian is given by
\begin{eqnarray}
e^{-1}{\cal L} &=& \varphi \Big (R + (D-1)(D-3) g^2\Big) + \varphi^3
\Big(-\ft14 {\cal F}_\2^2 + g^2(D-1)\Big) \cr 
&&+\varphi s^{ij}\left[\ft12 \bar \psi^i_\mu \Gamma^{\mu\nu\rho}
{\cal D}_\nu({\cal A}) \psi^j_\rho - \im
\sqrt{\ft{(D-2)\beta}{D-1}}\, \bar \lambda^i \Gamma^{\mu\nu} {\cal
D}_\mu({\cal A})\psi^j_\nu + \ft12 {\cal D}^\nu({\cal
A})(\bar\psi^i_\mu \Gamma^\mu \psi^j_\nu)\right]\cr 
&&+\varphi^2 t^{ij}\Big[\ft{{\rm i} \sqrt{\beta}}{16}\,
\bar\psi^i_\mu \Gamma^{\mu\nu\rho\sigma} \psi^j_\sigma - \ft{{\rm
i}\sqrt{\beta}}{8}\, \bar\psi^{i\nu}\psi^{j\rho} +
\sqrt{\ft{D-2}{16(D-1)}}\, \bar \psi^i_\mu
\Gamma^{\nu\rho}\Gamma^\mu\lambda^j \cr&&
\qquad\qquad+\sqrt{\ft{D-2}{16(D-1)}}\,\bar\psi^{i\nu}\Gamma^\rho
{\lambda}^j - \ft{{\rm i}\sqrt{\beta}(D-2)}{4(D-1)}
 \bar{\lambda}^i\Gamma^\nu  \Gamma^{\rho} \lambda^{j}
\Big] {\cal F}_{\nu\rho}\cr 
&&\varphi u^{ij}\left[  -\ft{\rm i\sqrt{\beta}}8 g \Big((D-1)\varphi
+ (D-2)\varphi^{-1}\Big) \bar \psi^i_\mu \Gamma^{\mu\nu} \psi_\nu^j
\right.\cr 
&&\qquad\qquad\left.- \ft12 \sqrt{(D-1)(D-2)}\, g \varphi
\bar\psi_\mu ^i \Gamma^\mu \lambda_j+\ft{\im}2 (D-2)\sqrt{\beta}\,g
\varphi \bar\lambda^i \lambda^j\right]\,,
\end{eqnarray}
where all the fermions are charged under the vector field ${\cal
A}_\1$, with the covariant derivative on fermions given by
\begin{equation}
{\cal D}_\mu({\cal A}) \xi^i = D_\mu \xi^i - \ft14\beta (D-3) {\cal
A}_\mu u^{ik} s^{jk} xi^j\,,
\end{equation}
for any spinor field $\xi^i$.  The pseudo-supersymmetric
transformation rules are given by
\begin{eqnarray}
{\delta}{\psi}^i_\mu &=& {\cal D}_\mu({\cal A})\epsilon^i -\ft{{\rm
i}\sqrt{\beta}}{4}\, {{} } t^{ij}{{} } s^{kj} {\varphi}^{-1} {\cal
F}_{\mu\rho} {\Gamma}^{\rho}{\epsilon}^k + \ft{{\rm
i}\sqrt{\beta}}{2}\,u^{ij}{{} } s^{kj}g\varphi{\Gamma}_\mu{{}
}{\epsilon}^k \,,\cr 
{\delta} {\lambda}^i &=& \ft{{\rm
i}\sqrt{\beta(D-1)}}{2\sqrt{(D-2)}}\Big[ \varphi^{-1}\Gamma^\mu
\partial_\mu\varphi\,{{} }{\epsilon}^i + \ft{{\rm i}\sqrt{\beta}}{4}\, {{} }
t^{ij}{{} } s^{kj}\varphi^{-1}{\cal F}_{\mu\nu}\,\Gamma^{\mu\nu}{{}
}{\epsilon}^k \cr 
&&\qquad\qquad-\ft{\im \sqrt{\beta}\, (D-3)}{2} u^{ij}{{} } s^{kj} g
(\varphi -\varphi^{-1})\Big]\,, \cr 
{\delta}\varphi &=& -\ft{{\rm i}}{4}\sqrt{\ft{(D-2)\beta}{D-1}}\,
{{} } s^{ij}\varphi\bar\lambda^i {{} }\epsilon^j\,,\qquad \delta
e^{a}_\mu= \ft14 s^{ij} \bar\psi_\mu^i  \Gamma^a {{} }\epsilon^j \,,
\cr 
{\delta} {\cal A}_\mu &=&\varphi^{-1}\, {{} } t^{ij}\,\Big[-\ft{{\rm
i}\sqrt{\beta}}{4}\, \bar\psi_\mu^i{\epsilon}^j  +
\ft{\beta\sqrt{D-2}}{4\sqrt{D-1}}\, \bar {\lambda}^i \Gamma_\mu {{}
}{\epsilon}^j \Big]\,. 
\end{eqnarray}
Thus we see that there is no term with any space-time derivative on
$\varphi$ in the full Lagrangian. The variation of $\varphi$ gives
rise to the quadratic solution of $\varphi$ and hence $\varphi$ can
be solved straightforwardly. Substituting the algebraic equation of
$\varphi$ back to the Lagrangian and pseudo-supersymmtric
transformation rules, we obtain the $f(R)$ theory of the gauged
Kaluza-Klein pseudo-supergravity. Note that the
pseudo-supersymmetric transformation rules on the fermions give rise
to the Killing spinor equations (\ref{dwks}) discussed earlier.

    Note that in this example, the Lagrangian in the $f(R)$ frame is
a polynomial of $\varphi$ with integer powers up to the cubic order
in all dimensions.  This is because the condition for
pseudo-supergravity is much less than supergravities for which
examples with integer powers are rare and dimensional dependent.

   To be self-contained, it is necessary that we present the
$\Gamma$-matrix and fermion conventions. We adopt exactly the same
convention given in \cite{lpwpseudo,llwpseudo}, which follows the
convention of \cite{vanp}.  We present the convention in Table 1. In
addition to the $\Gamma$-matrix symmetries and spinor
reperesentations in diverse dimensions, we also present the
$s^{ij}$, $t^{ij}$ and $u^{ij}$ that appear in the construction.

\bigskip\bigskip
\centerline{
\begin{tabular}{|c|c|c||c|c||c|c|c|}\hline
 $D$ mod 8 & $C\Gamma^{(0)}$ & $C\Gamma^{(1)}$ & Spinor &$\beta$
& $s^{ij}$ & $t^{ij}$ & $u^{ij}$\\ \hline\hline 
0&S&S&M&$+1$&$\delta^{ij}$&$\delta^{ij}$&$\varepsilon^{ij}$ \\
&S&A&S-M&$-1$&$\varepsilon^{ij}$&$\delta^{ij}$&$\varepsilon^{ij}$\\
\hline 
1&S&S&M&$+1$&$\delta^{ij}$&$\delta^{ij}$&$\varepsilon^{ij}$\\
\hline 
2&S&S&M&$+1$&$\delta^{ij}$&$\delta^{ij}$&$\varepsilon^{ij}$\\
&A&S&M&$-1$&$\delta^{ij}$&$\varepsilon^{ij}$&$\delta^{ij}$\\
\hline 
3&A&S&M&$-1$&$\delta^{ij}$&$\varepsilon^{ij}$&$\delta^{ij}$\\ \hline
4&A&S&M&$-1$&$\delta^{ij}$&$\varepsilon^{ij}$&$\delta^{ij}$\\
&A&A&S-M&$+1$&$\varepsilon^{ij}$&$\varepsilon^{ij}$&$\delta^{ij}$\\
\hline 
5&A&A&S-M&$+1$&$\varepsilon^{ij}$&$\varepsilon^{ij}$&$\delta^{ij}$\\
\hline 
6&A&A&S-M&$+1$&$\varepsilon^{ij}$&$\varepsilon^{ij}$&$\delta^{ij}$\\
&S&A&S-M&$-1$&$\varepsilon^{ij}$&$\delta^{ij}$&$\varepsilon^{ij}$\\
\hline 
7&S&A&S-M&$-1$&$\varepsilon^{ij}$&$\delta^{ij}$&$\varepsilon^{ij}$\\
\hline
\end{tabular}}
\bigskip

\begin{center}
Table 1: $\Gamma$-matrix symmetries and spinor reperesentations in
diverse dimensions. The quantities $s^{ij}$,$t^{ij}$ and $u^{ij}$
that appear in the Lagrangian take either $\delta^{ij}$ or
$\varepsilon^{ij}$.
\end{center}

The quantities $(s,t,u)$ satisify the following identities
\begin{equation}
s^{ij}s^{ik}=\delta^{jk}\,,\qquad s^{jk}t^{jl}s^{lm}=t^{km}\,,\qquad
s^{jk}t^{jl}=t^{kj}s^{lj}\,,\qquad s^{kj}t^{jl}=t^{kj}s^{jl}\,.
\end{equation}
Note that in the above, $s$ and $t$ can interchange, and each can
interchange with $u$ and the identities still hold.  We also have
the following important identities
\begin{equation}
s^{ji}t^{jk} s^{mk}t^{ml}=\beta\, \delta^{il}\,,\qquad
s^{ji}u^{jk}s^{mk}u^{ml}=-\beta\, \delta^{il}\,,\qquad s^{ji}u^{jk}
s^{lk}t^{lm} = \gamma \beta\,\varepsilon^{im}\,,
\end{equation}
where
\begin{equation}
\gamma =\Big\{ \begin{array}{cc} +1\,, & {\rm if~~}
t^{ij}=\delta^{ij} \,,\cr -1\,, & {\rm if~~} t^{ij}=\varepsilon^{ij}
\,.
\end{array}\label{gammavalue}
\end{equation}

\section{A general class of $f(R)$ pseud-supergravities}

It was shown \cite{llwfr} that there exist a subclass of $f(R)$
gravities that admit Killing spinor equations
\begin{equation}
D_\mu \epsilon + W\Gamma_\mu \epsilon=0\,,\qquad \Big(\Gamma^\mu
\nabla_\mu F + U\Big)\epsilon=0\,,
\end{equation}
where
\begin{equation}
U=-\fft{(4D(D-1)W^2 + R)f''(R)}{4(D-1)W'}\,.
\end{equation}
The function $f$ satisfies the following second-order linear
differential equation
\begin{equation}
f'' -\fft{\Big(4(D-1)(D-2) W^2+ R\Big)
W'}{\Big(4D(D-1)W^2+R\Big)W}\,f' +
\fft{W'}{\Big(4D(D-1)W^2+R\Big)W}\,f=0\,.\label{feom}
\end{equation}
As in the previous examples \cite{lpwpseudo,llwpseudo}, we find that
the existence of the Killing spinor equations implies that one can
performal pseudo-supersymmetric extension of these $f(R)$ gravities.
Introducing pseudo-gravitino and dilatino fields, we find that the
full Lagrangian is given by
\begin{eqnarray}
e^{-1}{\cal L} &=& f(R) + F(R)s^{ij}\left[\ft12 \bar \psi^i_\mu
\Gamma^{\mu\nu\rho}D_\nu \psi^j_\rho - \im
\sqrt{\ft{(D-2)\beta}{D-1}}\, \bar \lambda^i \Gamma^{\mu\nu}
D_\mu\psi^j_\nu + \ft12 D^\nu(\bar\psi^i_\mu \Gamma^\mu
\psi^j_\nu)\right]\cr 
&&+ \varphi u^{ij}\left[ -\ft14  \Big(2(D-2)W + U\Big) \bar
\psi^i_\mu \Gamma^{\mu\nu} \psi_\nu^j +\ft{\im
\sqrt{(D-1)(D-2)}}{\sqrt{\beta}}\, W  \bar\psi_\mu ^i \Gamma^\mu
\lambda_j \right.\cr &&\left.+ \ft12 \Big(-\ft{U'}{f''} +\ft{1}{D-2}
U + 2 W\Big) \bar \lambda^i\lambda^j\right]\,.\label{pseudofrlag}
\end{eqnarray}
The pseudo-supersymmetric transformation rules are given by
\begin{eqnarray}
{\delta}{\psi}^i_\mu &=& {\mathcal D}_\mu\epsilon^i+ \, u^{ij}{{} }
s^{kj} W {\Gamma}_\mu{{} }{\epsilon}^k \,,\cr 
{\delta} {\lambda}^i &=& \ft{{\rm
i}\sqrt{\beta(D-1)}}{2\sqrt{(D-2)}}\,F^{-1}\Big[\Gamma^\mu
\partial_\mu F\,{{} }{\epsilon}^i +U
u^{ij}{{} } s^{kj}{\epsilon}^k\Big]\,,\cr 
\delta e^{a}_\mu &=& \ft14 s^{ij} \bar\psi_\mu^i  \Gamma^a {{}
}\epsilon^j \,.
\end{eqnarray}
It can be shown that the Lagrangian is invariant under the
pseudo-supersymmetric transformation rules up to the quadratic order
in fermions. Note that in dimensions where $s^{ij}=u^{ij}$, the $i$
and $j$ indices in fermions can be suppressed.  The vanishing of the
pseudo-supersymmetric variation on $\psi$ and $\lambda$ gives rise
to precisely two Killing spinor equations obtained in \cite{llwfr}.

    It is worth observing that in the $f(R)$ frame, the dilatino
$\lambda$ is also an auxiliary field.  The variation of the
Lagrangian (\ref{pseudofrlag}) with respect to $\lambda$ gives an
algebraic equation for $\lambda$. Substituting this $\lambda$ back
to the Lagrangian give an $f(R)$ pseudo-supergravity involving only
the metric and gravitino.

\section{Conclusions and discussions}

   In this paper, we have constructed the $f(R)$ formalism of
ten-dimensional supergravities by performing a conformal
transformation and casting the theories in the $f(R)$ frame.  The
characteristics of the $f(R)$ frame is that the dilaton scalar
becomes auxiliary and hence can be integrated out.  In ten
dimensions, the $f(R)$ frame coincides with that of M-theory,
D2-branes or NS-NS 5-branes.  The process of integrating out the
auxiliary dilaton to get $f(R)$ gravity is analogous to integrate
out the auxiliary tensor fields in the Polyakov string action to
obtain the Nambu-Goto action.  When the fermionic sector are
included, we show up to quadratic order in fermions that the dilaton
remains auxiliary. The conclusion holds at the full (quartic)
fermionic orders since there is no derivative at this order.
Furthermore, the dilatino also becomes auxiliary and can be
integrated out. Using the same technique, we also constructed $f(R)$
theories of some $D=5$ and $D=4$ gauged supergravities and large
classes of pseudo-supergravities. We obtain many examples of BPS
$p$-brane and wormhole solutions in the $f(R)$ theories and analyze
their properties.

     There are two important issues to address at this stage.  The
first is whether the $f(R)$ supergravities are equivalent to the
corresponding usual supergravities in the Einstein frame.  The
second is which formalism is more natural.  To answer the first
question, we note that the main difference of the two formalism is
the conformal transformation.  If the conformal transformation is
non-singular, the two theories are clearly equivalent.  However,
when the conformal transformation becomes singular, the solution
spaces of the two theories become inequivalent.  This is because in
general relativity, a solution is not only characterized by the
local form, but also determined by the global structure.  It clear
that the local form of a solution in $f(R)$ supergravity can be
transformed in general to that of the corresponding supergravity,
the well-defined global structure of the solution in $f(R)$
supergravity can be destroyed in such a transformation if the
conformal factor is singular.  In this paper, we presented many
examples of well-defined solutions in $f(R)$ supergravities that
became badly behaved in the Einstein frame. In fact in terms of the
number of smooth solutions, the Einstein frame scores the worst.

     In ten dimensional supergravities, there is a large collection
of $p$-branes, and many physical properties of various $p$-brane
frames were discussed in \cite{dklsoliton}.  A distinguishing
feature we observed about the $f(R)$ frame is that the dilaton in
this frame becomes manifestly auxiliary.  The $f(R)$ frame coincides
with that of M-theory, D2-branes and NS-NS 5-branes.  The physical
significance of having an auxiliary coupling field for the D2-brane
and NS-NS 5-brane theories requires further investigation.  It could
be a consequence that the theories are intrinsically
non-perturbative and hence there is no such a field that gives rise
to the perturbrative expansion. It is also worth investigating
whether this is simply the artefact of lower-energy effective
action, or whether it persists when higher-order curvature
invariants are included.

    To address the second question, we note that while all
supergravities with a dilaton can be cast into the $f(R)$ frame,
many of them appear to have irrational powers of the scalar field.
For example, the bosonic Lagrangian of ${\cal N}=1$, $D=7$ gauged
supergravity in the $f(R)$ frame becomes
\begin{eqnarray}
e^{-1} {\cal L}_7 &=& \varphi R - g^2 \varphi^{\fft75}\Big( \ft14
\varphi^{-\fft{8\sqrt6}{5}} -2 \varphi^{-\fft{3\sqrt6}{5}} -2
\varphi^{\fft{2\sqrt6}{5}}\Big) \cr 
&&-\ft1{48} \varphi^{-\fft15 +\fft{4\sqrt6}{5}} F_\4^2 - \ft14
\varphi^{-\fft45 +\fft{2\sqrt6}{5}} (F_\2^i)^2 + e^{-1}{\cal
L}_{FFA}
\end{eqnarray}
Although the scalar $\varphi$ is auxiliary in this case, there is no
simple way of integrating out this field.  In ten dimensions,
however, in all supergravities, the scalar $\varphi$ in the $f(R)$
frame couples to the form fields with integer power.  We find that
this is because that the $f(R)$ frame is closely related to the
Kaluza-Klein $S^1$ reduction.  It turns out that the $f(R)$ frame is
the same as the $(D+1)$-dimensional frame without scaling in the
Kaluza-Klein reduction.  In other words, the breathing mode in the
Kaluza-Klein circle reduction is an auxiliary field and it becomes
manifest if we expressed the lower-dimensional theory in the
$(D+1)$-dimensional frame.  We demonstrate in the appendix that this
is true even when the higher-order Gauss-Bonnet term is considered.
However, a generic higher-order curvature term can introduce
derivatives on the breathing mode; nevertheless, the the
lower-dimensional theory still appears to be simplest in the $f(R)$
frame.

      Thus from the circle reduction of M-theory to $D=10$
supergravity, the $f(R)$ frame can be easily understood as the
M-theory frame.  It can also be easily understood why it also
coincides with that of D2-branes and NS-NS 5-branes.  From the
M-theory point of view, the $f(R)$ frame is clearly a natural frame
to work with. However, the question about which frame is more
natural to study supergravity can only be addressed when
higher-order curvature terms are included in the theory.  This is
because when higher-order curvature terms are involved, a conformal
transformation can dramatically increase the complexity of the
action.  For example, heterotic supergravity in $D=10$ is the
simplest in the string frame \cite{berderoo2}.  This suggests that
type IIA supergravity with $\alpha'$ correction is best constructed
in the string frame.  On the other hand, from our discussion on
$S^1$ reduction in the appendix, if we were simply to obtain type
IIA supergravity from $D=11$ with higher-order curvature invariants,
it is more natural to perform the reduction in the $f(R)$ frame.
This paradox requires further investigation.  It is related to the
fact that ten-dimensional supergravities play two roles in string
and M-theory. One is that they are the low-energy effective theories
of string; the other is that they are related to M-theory through
the dimensional compactification.

The $f(R)$ theories of supergravities can be viewed as $f(R)$
gravities coupled to matters including the form fields and also
fermions.  However, in the traditional $f(R)$ theories with matters,
the matter fields are typically coupled to the metric, whilst in
$f(R)$ supergravities, they are coupled not only to the metric, but
also to the curvature.  This enlarges the possibility of
constructing $f(R)$ theories coupled with matters.

      To conclude, we find that there exists the $f(R)$ formalism for
supergravities.  Our explicit construction of $f(R)$ supergravities
allows us to find new BPS solutions and also leads to new questions
that are worth further investigation.

\section*{Acknowledgement}

We are grateful to Chris Pope for useful discussions, and grateful
to KITPC, Beijing, for hospitality during the part of this work. Liu
is supported in part by the National Science Foundation of China
(10875103, 11135006) and National Basic Research Program of China
(2010CB833000). L\"u is supported in part by the NSFC grant
11175269.

\appendix

\section{KK circle reduction without the conformal scaling}

Kaluza-Klein circle reduction has been well-studied. It was
typically done with the lower-dimensional metric properly scaled by
the breathing mode such that the theory in lower dimensions is in
the Einstein frame.  In this picture, the breathing mode is a
dynamical field and cannot be solved algebraically.  Here we present
the Kaluza-Klein circle reduction from $(D+1)$ dimensions to $D$
dimensions, with the $D$-dimensional metric written in the same
frame as that in $(D+1)$ dimensions.  The metric ansatz is given by
\begin{equation}
d\hat s^2_{D+1} = ds_D^2 + \varphi^2 (dz + {\cal A}_\1)^2\,.
\label{frred}
\end{equation}
The lower-dimensional metric,  the breathing mode $\varphi$ and the
Kaluza-Klein vector ${\cal A}_\1$ are all independent of the
coordinate $z$. The natural choice of the viebein and their inverse
are given by
\begin{eqnarray}
&&\hat e^a=e^a\,,\qquad e^{\bar{z}} = \varphi (dz + {\cal
A}_\1)\,,\cr 
&&\hat E_a=\nabla_a - A_a \partial_z\,,\qquad \hat E_{\bar z}
=\varphi^{-1}\partial_z\,,
\end{eqnarray}
where the $(D+1)$-dimensional world and tangent indices are split to
$\hat\mu=(\mu, z)$ and $\hat a=(a,\bar{z})$ respectively.  Note that
we have $\nabla_a=E_a^\mu \nabla_\mu$ and ${\cal A}_a=E_a^\mu {\cal
A}_\mu$. The components of the spin-connection $\omega^{ab}\equiv
\omega_c{}^{ab} e^c$ are given by
\begin{equation}
\hat\omega_{cab}=\omega_{cab}\,,\qquad \omega_{\bar z a
b}=-\ft12\varphi {\cal F}_{ab}\,,\qquad \omega_{b\bar z
a}=\ft12\varphi {\cal F}_{ab}\,,\qquad \omega_{\bar z\bar z
a}=\varphi^{-1} \nabla_a\varphi\,.
\end{equation}
Here we have ${\cal F}_\2=d{\cal A}_\1$.  Since we are interested in
also the reduction of the higher-order curvature terms, we present
the reduction on the Riemann tensors as well as the Ricci tensors.
The independent non-vanishing components of the Riemann tensor are
given by
\begin{eqnarray}
\hat R^{ab}{}_{cd}&=&R^{ab}{}_{cd} -\ft12\varphi^2\big( {\cal
F}^{a}{}_{b} {\cal F}_{cd}- {\cal F}^{a}{}_{[c} {\cal
F}_{d]b}\big)\,, \cr 
\hat R_{abc{\bar z}}&=&\hat R_{c{\bar z}ab} =
\nabla_{[a}\big(\varphi\, {\cal F}{}_{b]c}\big) -
\nabla_c\phi\,{\cal F}_{ab} \,, \cr 
\hat R^{{\bar z}}_{~a{\bar z} b}&=&-\varphi^{-1}\nabla_b
\nabla_{a}\varphi +\ft14\varphi^2\, {\cal F}_{cb} {\cal F}^{c}{}_{a}
\,.
\end{eqnarray}
The independent components of the Ricci-tensor are
\begin{eqnarray}
\hat R_{ab}&=&R_{ab} -\varphi^{-1}\nabla_b \nabla_{a}\varphi
-\ft12\varphi^2 {\cal F}_{cb} {\cal F}^{c}{}_{a} \,, \cr 
\hat R_{a{\bar z}}&=& -\ft12\varphi^{-2} \nabla_{c}\big(\varphi^3
{\cal F}^c{}_{a}\big)\,, \cr 
\hat R_{{\bar z}{\bar z}}&=&-\varphi^{-1}\Box\varphi
+\ft14\varphi^2\, {\cal F}_{(2)}^2\,.
\end{eqnarray}
The Ricci scalar is
\begin{eqnarray}
\hat R&=&R-\ft14\varphi^2 {\cal F}_{(2)}^2 -2\varphi^{-1}\Box
\varphi \,.
\end{eqnarray}
Thus we have
\begin{equation}
{\cal L}_{D+1}=\sqrt{-\hat g}\,\hat R\qquad \longrightarrow \qquad
{\cal L}_D=\sqrt{-g}\, \Big(\varphi R - \ft14\varphi^3 {\cal F}_\2^2
-\Box\varphi\Big)\,.
\end{equation}
Note that the last term is a total derivative and hence can be
dropped.  It is easy to verify that the equations of motion derived
from the $D$-dimensional Lagrangian satisfy the Einstein equations
in $(D+1)$ dimensions. In this reduction scheme, the breathing mode
$\varphi$ in $D$ dimensions is an auxiliary field, and can be
integrated out. In other words, the $D$-dimensional theory is in the
``$f(R)$ frame.''

     If $(D+1)$-dimensional theory has higher-order curvature
terms, it becomes more apparent that the reduction ansatz
(\ref{frred}) is more natural in that it gives rise to the simplest
form in the lower-dimensional theory.  For a generic higher-order
curvature term, the breathing mode can acquire derivatives and cease
to be auxiliary.  This is analogous to off-shell supergravities
where the auxiliary fields acquire dynamics when higher-order
supersymmetric invariants are introduced.  However, if the
higher-order curvature terms have special properties such as being
topological, the breathing mode may remain auxiliary.  Let us
demonstrate this by reducing Einstein gravity with the Gauss-Bonnet
term, namely
\begin{equation}
\hat e^{-1} {\cal L}_{D+1} = \hat R - \Lambda_0 + \alpha \hat E_{\rm
GB}\,,\label{dp1gb}
\end{equation}
where
\begin{equation}
\hat E_{\rm GB} = \hat R^2 - 4 \hat R^{\hat\mu\hat\nu} \hat
R_{\hat\mu\hat \nu} + \hat R^{\hat \mu\hat\nu\hat \rho \hat\sigma}
\hat R_{\hat \mu\hat\nu \hat\rho \hat\sigma}\,.
\end{equation}
For simplicity and demonstrating the point, let us for now set
${\cal A}_\1=0$.  The reduced Gauss-Bonnet term is given by
\begin{equation}
\hat E_{\rm GB}= E_{\rm GB} + 8\varphi^{-1} \nabla_a\nabla_b \varphi
R^{ab} - 4\varphi^{-1} \Box \varphi R\,.
\end{equation}
It follows that for ${\cal A}_\1=0$, we have
\begin{equation}
\sqrt{-\hat g} (\hat R -\Lambda_0+ \alpha \hat E_{\rm GB}) =
\sqrt{-g}\, \varphi ( R -\Lambda_0 + \alpha E_{\rm GB}) +
\hbox{total derivative terms}\,.
\end{equation}
In other words, there is no term involving a derivative on
$\varphi$. This is true even when the Kaluza-Klein vector is turned
on.  The full $S^1$ reduction of the Lagrangian (\ref{dp1gb}) on
$S^1$ with the reduction (\ref{frred}) is given by
\begin{eqnarray}
e^{-1} {\cal L}_D &=& \varphi (R -\Lambda_0 + \alpha E_{\rm GB})
-\ft14 \varphi^3 {\cal F}_\2^2 +\alpha \varphi^3\Big(- R^{abcd}
({\cal F}_{ab} {\cal F}_{cd} - {\cal F}_{ac} {\cal F}_{db})\cr 
&&+2\nabla_a {\cal F}_{bc} \nabla^a {\cal F}^{bc} + \ft13 \Box {\cal
F}_\2^2 -2\nabla_a\nabla_b ({\cal F}^2)^{ab} + 2 \nabla_b {\cal
F}^{ba} \nabla^c {\cal F}_{ca}\cr 
&&-\ft{10}3 \nabla_a{\cal F}_{bc} \nabla^{b} {\cal F}^{ac} +4 F^{ab}
\nabla_a\nabla^c {\cal F}_{cb} - \ft43 {\cal F}^{ab}
\nabla^c\nabla_b {\cal F}_{ac}\Big)\cr 
&&+\alpha\varphi^5 \Big(\ft14 ({\cal F}_\2^2)^2 + \ft12 ({\cal
F}_\2^2)^{ab} ({\cal F}_\2^2)_{ab} \Big)\,,
\end{eqnarray}
where $({\cal F}_\2^2)^{ab} = {\cal F}^{a}{}_c {\cal F}^{bc}$.  Thus
we see that the breathing mode $\varphi$ is an auxiliary field and
can be integrated out.

\section{A general class of charged black hole solutions}

We find that the Lagrangian (\ref{gaugegenlag}) admits the following
charged black hole solution
\begin{eqnarray}
ds_D^2&=&-{\cal H}^{-\fft{D-3}{D-2}} H^{-\fft{D-1}{D-2}}\, h\, dt^2
+ {\cal H}^{\fft{1}{D-2}} H^{\fft{D-1}{(D-2)(D-3)}}
\Big(\fft{dr^2}{h} + r^2 d\Omega_{D-2,k}^2\Big)\,,\cr 
{\cal F}_\2&=& \sqrt{k}\coth(\sqrt{k}\,\tilde\delta)\,dt\wedge
d{\cal H}^{-1}\,,\quad F_\2=\sqrt{\fft{D-1}{D-3}}\,\sqrt{k}\,
\coth(\sqrt{k}\,\delta)\,dt\wedge dH^{-1}\,,\cr 
h &=& k -\fft{\mu}{r^{D-3}} + g^2 r^2 {\cal H}
H^{\fft{D-1}{D-3}}\,,\qquad e^{\phi} = \Big(\fft{\cal
H}{H}\Big)^{\sqrt{\fft{D-1}{2(D-2)}}}\,,\cr 
{\cal H} &=& 1 +
\fft{\mu\sinh^2(\sqrt{k}\,\tilde\delta)}{k\,r^{D-3}}\,,\qquad H=1 +
\fft{\mu\sinh^2(\sqrt{k}\,\delta)}{k\,r^{D-3}}\,.\label{cbh0}
\end{eqnarray}
Here the parameter $k$ can be 1, 0, or $-1$, corresponding to the
cases where the foliation in the transverse space have the metric
$d\Omega_{D-2,k}^2$ on the unit $S^{D-2}$, $T^{D-2}$ or $H^{D-2}$,
where $H^{D-2}$ denotes the unit hyperbolic $(D-2)$-space of
constant negative curvature.  Note that the solution has a smooth
limit for $k=0$ and it remains real when $k=-1$.  The horizon of the
black hole is located at the largest root of the function $h$.  When
$h$ has a double root, the solution becomes extremal. In $D=5$, the
solution is a special case of the $U(1)^3$ charged solutions in
five-dimensional gauged supergravity \cite{bcs2}.  In $D=4$, it is a
special case of the $U(1)^4$ charged solutions in four-dimensional
gauged supergravity \cite{duffliu}. (See also \cite{sabra}.)  Note
that when $F_\2=0$, the single-charg Kerr-AdS black holes of
(\ref{gaugedKKlag}) was constructed in \cite{wu}.

     There is another limit one can take for $k=1$ and 0 cases.  We
can send $\mu\rightarrow 0$, $\delta\rightarrow \infty$ and
$\tilde\delta\rightarrow \infty$, but with $q=\mu \sinh^2\delta$ and
$\tilde q=\mu \sinh^2\tilde\delta$ fixed.  In the case of $D=4,5$,
this is a BPS limit and the resulting solutions preserve a fraction
of the supersymmetry.  For $k=0$, this limit implies that the
charges are set to zero, and we obtain a domain wall solution.  For
$k=1$, this limit gives rise to the solution (\ref{cbh1}).  Note
that such a limit cannot be taken for the $k=-1$ case.  The
properties of the solutions were discussed in sections 3.2, 6.1,
6.2. In particular, it was shown that the singular solutions with
$q\ne0$ and $\tilde q=0$ becomes smooth charged wormholes in the
corresponding $f(R)$ theories.

\end{document}